\documentclass[sn-apa,Numbered]{sn-jnl}

\usepackage{graphicx}%
\usepackage{multirow}%
\usepackage{amsmath,amssymb,amsfonts}%
\usepackage{amsthm}%
\usepackage{mathrsfs}%
\usepackage[title]{appendix}%
\usepackage{xcolor}%
\usepackage{textcomp}%
\usepackage{manyfoot}%
\usepackage{booktabs}%
\usepackage{listings}%
\usepackage{tikz}
\usepackage{comment}
\usepackage{bbm}
\usepackage{xspace}
\usepackage{float}

\newtheorem{prop}{Proposition}%
\newtheorem{example}{Example}%
\newtheorem{definition}{Definition}%

\newcommand*{\task}[4]{
	\draw[rounded corners] (#3-#2,#4+0.1) rectangle (#3,#4+0.6);
	\draw (#3-#2/2,#4+0.35) node {#1};
}

\newcommand{\gcheck}{{\color{green}\cmark}}%
\newcommand{\rcross}{{\color{red}\xmark}}%

\newtheorem{lemma}{Lemma}
\newtheorem{corol}{Corollary}

\newcommand{\fixme}[1]{{\color{blue} [FixMe] {#1}}}
\newcommand{\sigmaD}{$\Sigma Dev$\xspace}
\newcommand{\sigmaE}{$\Sigma $E\xspace}
\newcommand{\sigmaT}{$\Sigma $T\xspace}
\newcommand{\sigmaU}{$\Sigma $U\xspace}

\newcommand{\sigmaUgraph}{\textsc{$\Sigma U$-graph}\xspace}
\newcommand{\sigmaDgraph}{\textsc{$\Sigma Dev$-graph}\xspace}
\newcommand{\sigmaTgraph}{\textsc{$\Sigma T$-graph}\xspace}
\newcommand{\sigmaUinf}{\textsc{$\Sigma U$-inferred}\xspace}
\newcommand{\sigmaDinf}{\textsc{$\Sigma Dev$-inferred}\xspace}
\newcommand{\sigmaTinf}{\textsc{$\Sigma T$-inferred}\xspace}

\newcommand{\sigmaTgraphproblem}{\textsc{$\Sigma T$-graph}\xspace problem\xspace }
\newcommand{\chainsSigmaU}{$(1|chains,p_j\!=\!1|\Sigma U_j)$\xspace}
\newcommand{\chainsSigmaT}{$(1|chains,p_j\!=\!1|\Sigma T_j)$\xspace}

\newcommand{\prefsched}{scheduling preferences\xspace}

\newcommand{\precedencegraphe}{precedence graph\xspace}
\newcommand{\precedenceinduite}{inferred precedences\xspace}

\newcommand{\critbinary}{Binary Criterion\xspace}
\newcommand{\critdistance}{Distance Criterion\xspace}

\begin{document}

\title[Collective schedules]{Ordering Collective Unit Tasks: from Scheduling to Computational Social Choice}


\author*{\fnm{Martin} \sur{Durand}}\email{martin.durand@lip6.fr}

\author{\fnm{Fanny} \sur{Pascual}}\email{fanny.pascual@lip6.fr}

\affil{\orgdiv{Sorbonne Université}, \orgname{LIP6, CNRS}, \orgaddress{\street{5 Place Jussieu}, \city{Paris}, \postcode{75005}, \country{France}}}

\abstract{We study the collective schedules problem, which  consists in computing a one machine schedule of a set of tasks, knowing that a set of individuals (also called voters)  have preferences regarding the order of the execution of the tasks. Our aim is to return a consensus schedule. We consider the setting in which all tasks have the same length -- such a schedule can therefore also be viewed as a ranking.  
We study two rules, one based on a distance criterion, and another one based one a binary criterion, and we show that these rules extend classic scheduling criteria. We also consider time constraints and precedence constraints between the tasks, and focus on two cases: the preferences of the voters fulfill these constraints, or they do not fulfill these constraints (but the collective schedule should fulfill them). In each case, either we show that the problem is NP-hard, or we provide a polynomial time algorithm which solves it. We also provide an analysis of a heuristic, which appears to be a 2 approximation of the Spearman's rule.        
}




\maketitle

 The collective schedules problem~\citep{pascual2018collective,durand2022collective} consists in scheduling a set of $n$ tasks shared by $v$ individuals, also called  voters. 
 The tasks may represent talks of a conference that will be done in a same room, works to be done sequentially by co-workers, or events that will occur in one of the weekly meetings of an association. Each voter has  his or her own preferences regarding the order in which theses tasks will be executed. We consider two models. In the first one, introduced by~\cite{pascual2018collective} and that we will call  \emph{Order Preferences}, each voter gives his or her preferred order -- a permutation of the tasks. In the second model, that we introduce in this paper, and that we call \emph{Interval Preferences}, each voter gives for each task the interval in which he or she would like the task to be done. In this paper, we focus on situations in which all the tasks have the same duration (a time slot per task). Our aim is, given the preferences of each individual, to compute a good compromise schedule of the $n$ tasks. 

Note here that, since the tasks are unit tasks, a schedule of the $n$ tasks can be seen as a ranking of $n$ tasks (or candidates).  In this paper, we will use several   concepts -- such as precedence constraints -- from the scheduling field: we will therefore use the term schedule and not ranking for a permutation of the $n$ tasks. Note however that several results of this paper are interesting not only in the context of scheduling but also in the context of ranking candidates.
 
 The dissatisfaction of a voter if the returned schedule  is  schedule $S$ is measured thanks to two  families of criteria, coming from the scheduling theory field. One is a binary criterion, which says that a voter is satisfied if a task is not scheduled too late (or not too  early) in $S$ with respect to the preferences of the voters (expressed as a order preferences or interval preferences). The other family of criteria is a distance criterion, which says that the closer the returned schedule is to the voter's preferences, the more satisfied a voter is.

We measure the quality of a compromise schedule $S$ for all the voters by summing up the sum of the dissatisfaction of the voters for schedule $S$. This sum, divided by $v$, represents the average dissatisfaction of a voter with solution $S$. We focus on an utilitarian criterion: our aim is to compute a schedule with the smallest sum of dissatisfaction. 

\medskip

\noindent {\bf An assignment problem.} Without additional constraints, this problem can be solved polynomially, both for Order and Interval Preferences, as it is an assignment problem. Indeed, the returned schedule being a permutation of the $n$ tasks, we know that there will be $n$ time slots, between 0 and $n$, one for each task. We create a complete bipartite graph with the tasks on the left and the time slots on the right. For each couple (task $t$, time slot $s$), the cost of the edge ($t,s$) is the sum of the dissatisfaction  caused by task $t$ to all the voters if $t$ is scheduled at time slot $s$. Therefore, a schedule that minimizes the total dissatisfaction corresponds to a minimum cost matching in such a graph. The graph can be built  in $O(v n^2)$, and a minimum cost matching can be found with Hungarian algorithm in $O(n^3)$~\citep{Tomizawa71,EdmondsK72}, leading to a $O(v n^2 + n^3)$ algorithm.

\medskip

\noindent {\bf Additional constraints.} Our aim is to study this problem by adding the main constraints in scheduling: time constraints and precedence constraints. Time constraints mean that to each task is associated a release date and a due date (or deadline), and that in the returned schedule each task should be scheduled between its release date and its deadline. Precedence constraints mean that there is a precedence graph of the $n$ tasks:  if there is an edge from task $i$ to task $j$ in this graph, this means that in the returned schedule task $i$ should be scheduled before task $j$. We will study both the case where these constraints are  imposed, and the case where they are inferred from the preferences of the voters. Before presenting our results and the map of the paper, we review related works. 

\medskip
\noindent {\bf Related work. }
Our work is at the boundary between computational social choice~\citep{brandt2016handbook} and scheduling~\citep{brucker1999scheduling}, two major domains in operational research. 

As mentioned above, the collective schedule problem generalizes the collective ranking problem, which is an active field in computational social choice (see e.g. ~\citep{DworkKNS01,SkowronLBPE17,CelisSV18,Singh2018,Biega2018,Geyik2019,Asudeh2019,Narasimhan2020}). In this field, authors often design rules (i.e. algorithms) which return fair rankings, and they  often focus on fairness in the beginning of the rankings. If the items (or candidates)  to be ranked are recommendations (of restaurants, web pages, etc.) for users, the beginning of the ranking is indeed probably the most important part. Note that this does not  hold for our problem since all the planned tasks will be executed -- only their order matters.  This means that rules designed for the collective ranking problem are not suitable not only because they do not consider duration for the items, but also because they focus on the beginning of the ranking. This also means that the rules that we will study can be relevant for consensus ranking problems where the whole ranking is of interest.

The collective schedule problem has been introduced in~\citep{pascual2018collective} for Order Preferences and when tasks have different processing times (lengths). In this paper, the authors introduced a weighted variant of the Condorcet principle, called the PTA Condorcet principle (where PTA stands for ``Processing Time Aware''), and they adapted  previously known  Condorcet consistent rules when tasks have different processing times. They also introduced a new rule, which computes a schedule which minimizes the sum of the tardiness of tasks between the preferred schedules of the voters and the schedule which is returned. They show that the optimization problem solved by this rule  is NP-hard but that it can be solved for reasonable size instances with a linear program.

~\cite{durand2022collective}  also study the collective schedule problem with Order Preferences and when the tasks have different lengths. They propose an axiomatic study of three rules: a generalization of the Kemeny rule, and the two rules which minimizes the sum of the deviations of the tasks between the preferred schedules and the returned schedule (rule \sigmaD, optimal for the total deviation criterion), and the sum of the tardiness  (rule \sigmaT, optimal for the total tardiness criterion).
They show that these rules solve NP-hard
problems, but that it is possible to solve optimally these problems for reasonable size instances.

Multi agent scheduling problems also aim at returning  consensus schedules, but they focus on cases where (usually two) agents own their \emph{own} tasks, that are scheduled on shared machines: the aim is to find a Pareto-optimal and/or a fair schedule of the tasks of the agents, each agent being interested by her own tasks only~\citep{SauleT09,agnetis2014}. 

Last but not least,  the work of~\cite{ElkindKT22} focus  on the assignment of shared unit size tasks to specific unit size time slots. This latter work focus on fairness notions extended from the multi-winner voting problem and does not use scheduling notions such as precedence constraints or time constraints.

\medskip
\noindent {\bf Our results and map of the paper. }
\begin{itemize}
    \item We first start by introducing notations in Section~\ref{sec:preliminaries}, as well as formal definition of the binary and distance criteria studied in this paper. As we will see, these criteria generalize the other criteria studied before (total  tardiness, and total deviation), and also allow us to model famous scheduling criteria, as the minimization of the total earliness of the tasks, or also the minimization of the number of late tasks. Rules that return optimal solutions of these criteria will be studied in the sequel.
    \item In Section~\ref{sec:EMD}, we focus on the 
algorithm which, in the Order Preference setting, computes the median start time of each task, and then schedules the tasks by increasing median start times (rule EMD -- for Earliest Median Date). We show that, interestingly, this rule returns a schedule which is a 2-approximation of the total tardiness criterion. 
\item We then focus in Section~\ref{sec:timeconstraints} on time constraints: we show that it is still possible to get an optimal solution in polynomial time with time constraints on the tasks. We focus on the rules optimizing the binary and distance criteria (without time constraints), as well as the EMD rule,  and we present an axiomatic study of these rules when time constraints are induced by the preferences of the voters (e.g. if all the voters schedule, in their preferred schedules, a task $t$ at time $X$, is this task $t$ necessarily started exactly at time $X$ in the returned schedule ? If task $t$ is always started after time $X$ in the preferred schedules, is it always scheduled after time $X$ in an optimal solution ?). 
\item In Section~\ref{sec:precedence}, we focus on  precedence constrains between the tasks. We show that the previously studied rules, which could be run in polynomial time without precedence constraints, can still be used (with an additional polynomial time step) when the precedence constraints are inferred by the preference of the voters.  On the contrary, we show that we have to solve NP-hard problem when the precedence constraints are not fulfilled by the preferred schedules of the voters. This is true both for the distance and the binary criterion, and in particular in the cases where we wish to minimize the total deviation, the total tardiness, or the number of late tasks. 
\item We conclude this paper in Section~\ref{sec:conclusion} by an overview of our results and a few research directions. 
\end{itemize}

\section{Preliminaries}
\label{sec:preliminaries}
\subsection{Definitions and notations}
\noindent {\bf Order Preferences and Interval Preferences. }

 We consider a set  $J=\{1, \dots, n\}$ of $n$ tasks of interest for a set $V=\{1, \dots, v\}$ of $v$ voters. Each task has a processing time of 1.  The preferences of voter $i$ are denoted by $\mathcal{V}_i$, and depend of the setting used. 

 In the \emph{Order Preferences} setting, each voter indicates its preferred schedule, as a permutation of the $n$ tasks (we do not consider idle times between the tasks). Therefore, $\mathcal{V}_i$ is the preferred schedule of voter $i$. We denote by $C_j(\mathcal{V}_i)$ the completion time of task $j$ in the preferred schedule of voter $i$. More generally, given a schedule $S$ of tasks of $J$, we denote by $C_j(S)$ the completion time of task $j$ in $S$.

 In the \emph{Interval Preferences} setting, each voter indicates for each task the interval in which he or she wishes to see the task scheduled. More precisely, for each task $j\in J$, voter $i\in V$ indicate a release date -- that will be denoted by  $r_j(\mathcal{V}_i)$  -- , and which means that voter $i$ would like task $j$ to be started at the soonest at time $r_j(\mathcal{V}_i)$. Likewise, voter $i\in V$ indicates a due date (also called deadline) -- that will be denoted by  $d_j(\mathcal{V}_i)$  --, and which means that voter $i$ would like task $j$ to be completed  at the latest at time $d_j(\mathcal{V}_i)$. Therefore, $\mathcal{V}_i$ is the set of the $n$ couples (release date, due date) that voter $i$ sets for the $n$ tasks. Note that this setting generalizes the Order Preferences settings, since it is possible for a voter to set for each task a release date (resp. a due date) equal to its start (resp. its completion time) in its preferred schedule. The only constraint we impose is that there exists a feasible schedule that fulfills the time constraint given by a voter (i.e. in which each task $j$ is scheduled in the interval $[r_j(\mathcal{V}_i),d_j(\mathcal{V}_i)]$).  
 
 In the sequel, we will penalize schedules in which tasks are scheduled out of the intervals given by the voters. The Interval Preferences setting allows  voters to express pretty precise preferences. Indeed, if a voter wants a task to be done before a given date $t$, and has no preference on the starting date of a task then she can indicate a release date of $0$ and a due date of $t$. If her only wish is that a task starts after a given time $t'$, then she can indicate a release date of $t'$ and a due date of $n$. Finally, if a voter wants a task to start exactly at time $t''$, then she can give a release date of $t''$ and a due date of $t''+1$. This flexibility in the preferences allow voters to express situations in which they have different expectations regarding the task, from having no interest in a task to wanting it to be completed exactly at a given time. 

\medskip
 Let us now present the two general objective functions that we will consider in this paper: the binary criterion, and the distance criterion.

\medskip
\noindent {\bf Binary criterion. }



The first criterion, called \emph{\critbinary}, measures whether a task is executed in the time interval indicated by a voter  or not (the penalty is $0$ if the tasks is scheduled in the desired interval, and is $1$ if the task is not scheduled in the desired interval). The dissatisfaction of voter $i\in V$ concerning task $j\in J$ is thus: 
$$
b_j(S,\mathcal{V}_i)=
\begin{cases}
1 & \text{if } C_j(S)>d_j(\mathcal{V}_i) \text{ or } C_j(S) \leq r_j(\mathcal{V}_i)\\
0 & \text{otherwise}
\end{cases}
$$

The dissatisfaction of a voter $i$ concerning a schedule $S$ with the binary criterion is then:  $$
B(S,\mathcal{V}_i)=\sum_{j \in J
} b_j(S, \mathcal{V}_i)
$$

\medskip
\noindent {\bf Distance criterion. }

The second criterion, called {\em \critdistance}, also does not count any penalty when a task is scheduled in its time interval, but otherwise it counts a penalty which expresses how far from its interval the task is.  
The dissatisfaction of voter $i$ concerning task $j$ in schedule $S$ is: $$
d_j(S,\mathcal{V}_i)=
\begin{cases}
C_j(S)-d_j(\mathcal{V}_i) & \text{if } C_j(S)>d_j(\mathcal{V}_i)\\
r_j(\mathcal{V}_i)-(C_j(S)-1) & \text{if } C_j(S) \leq r_j(\mathcal{V}_i)\\
0 & \text{otherwise}
\end{cases}
$$


The dissatisfaction of a voter $i$ concerning a schedule $S$ with the distance criterion is then: 
$$
D(S,\mathcal{V}_i)=\sum_{j \in 
J
} d_j(S, \mathcal{V}_i)
$$

\medskip
\noindent {\bf Aggregation function. } 

As said in the introduction, we will study the utilitarian utility function.  Our aim will be to minimize $\Sigma_{i\in\{1,\dots,v\}}B(S,\mathcal{V}_i)$ with the binary criterion, or $\Sigma_{i\in\{1,\dots,v\}}D(S,\mathcal{V}_i)$ with the distance criterion. The \emph{Binary Criterion rule} is an algorithm that returns a schedule minimizing binary criterion, while the \emph{Distance Criterion rule} is an algorithm that returns a schedule minimizing distance criterion.

\subsection{Generalization of classical scheduling criteria. }
\label{subsec:genralization}

The two above defined criteria generalize the main criteria already studied in the Order Preferences setting~\citep{pascual2018collective,durand2022collective}.
Assume indeed that voters have expressed their preferences using the Order Preferences setting (i.e. each voter indicates his or her preferred schedule). Let $P$ be the preference profile, i.e. the set of all the preferences expressed by the voters. 

\begin{itemize}
    \item {\em Total deviation}. The total deviation  of  a schedule $S$ is $Dev(S,P)= \Sigma_{i\in\{1,\dots,v\}}Dev(S, \mathcal{V}_i)$, where $Dev(S, \mathcal{V}_i)=\sum_{j \in {J}} |C_j(S)-C_j(\mathcal{V}_i)|$.         
        If our aim is to compute a schedule of minimal total deviation, as does rule \sigmaD~\citep{pascual2018collective,durand2022collective}, then we should use the Distance Criterion by setting the release date of task $j$ for voter $i$ at $C_j(\mathcal{V}_i) - 1$, and the due date  of task $j$ for voter $i$ at $C_j(\mathcal{V}_i)$. 
        \item {\em Total tardiness}. 
        The total tardiness of  a schedule $S$ is $T(S,P)= \Sigma_{i\in\{1,\dots,v\}}T(S, \mathcal{V}_i)$, where $T(S, \mathcal{V}_i)=\sum_{j \in {J}} \max(0, C_j(S)-C_j(\mathcal{V}_i))$.         
        If our aim is to compute a schedule of minimal total tardiness, as does rule \sigmaT~\citep{pascual2018collective,durand2022collective}, then we should use the Distance Criterion by setting the release date of task $j$ for voter $i$ at $0$, and the due date  of task $j$ for voter $i$ at $C_j(\mathcal{V}_i)$.

            \item {\em Total earliness}.   The total earliness of  a schedule $S$ is $E(S,P)= \Sigma_{i\in\{1,\dots,v\}}E(S, \mathcal{V}_i)$, where $E(S, \mathcal{V}_i)=\sum_{j \in {J}} \max(0, C_j(\mathcal{V}_i)-C_j(S))$. 
             If our aim is to minimize the total earliness, a classic  criterion in scheduling~\citep{brucker1999scheduling}, then we should use the Distance Criterion by setting the release date of task $j$ for voter $i$ at $C_j(\mathcal{V}_i)-1$, and the due date  of task $j$ for voter $i$ at $n$.
             \item {\em Total number of late tasks}. 
             The total number of late tasks of a schedule $S$  is $U(S,P)= \Sigma_{i\in\{1,\dots,v\}}U(S, \mathcal{V}_i)$, where $U(S, \mathcal{V}_i)$ is the number of tasks of ${J}$ such that $C_j(S) > C_j(\mathcal{V}_i)$ (such tasks are called \emph{late} tasks). 
             This a classic  criterion, denoted by \sigmaU (for ``Unit Penalties"), in scheduling~\citep{brucker1999scheduling}. We can solve this optimization problem by using the Binary Criterion by setting the release date of task $j$ for voter $i$ at $0$, and the due date  of task $j$ for voter $i$ at $C_j(\mathcal{V}_i)$.
             \item {\em Total number of tasks not well positioned}. If our aim is to maximize the number of tasks scheduled at the exact position given by the voters, then we should use the Binary Criterion by setting the release date of task $j$ for voter $i$ at $C_j(\mathcal{V}_i)-1$, and the due date  of task $j$ for voter $i$ at $C_j(\mathcal{V}_i)$.
             
\end{itemize}

In the next section, we introduce the EMD rule and show that it is a 2 approximation of the total tardiness (\sigmaT) and total earliness  (\sigmaE) criteria. 

\section{An analysis of the EMD rule}
\label{sec:EMD}

The EMD rule, introduced as a heuristic by~\cite{durand2022collective} in the Order Preferences setting, schedules the tasks by increasing median completion times, where the median time of a task $j$ is the median of the set $\{C_j(\mathcal{V}_1), \dots, C_j(\mathcal{V}_v)  \}$.  If several tasks have the same median completion time, any tie breaker mechanism can be used. 

It was shown  previously~\citep{pascual2018collective} that, for unit size tasks, and for any preference profile $P$  and any schedule $S$, we have $Dev(S,P)= 2 T(S,P)$, and thus that $T(S,P)=E(S,P)$ since  $Dev(S,P)= E(S,P) + T(S,P)$. Therefore, a $\alpha$-approximate algorithm\footnote{An $\alpha$-approximate algorithm, for a minimization problem (as are our criterion), returns, for each instance $I$, a solution (schedule) of cost at most $\alpha OPT(I)$, where $OPT(I)$ is the minimal cost that can be obtained on a solution of instance $I$.} for the total tardiness criterion will also be an  $\alpha$-approximate algorithm for the earliness criteria, and an  $\alpha$-approximate algorithm for the total deviation criterion.

We consider that we are in the Order Preferences setting. Before showing that EMD is 2-approximate for the total tardiness criterion (and thus also for the deviation and earliness criterion), we introduce a way to see the instance that will facilitate the analysis. 

\medskip

\noindent {\bf Breaking down the preference profile.} 
We ``break down" the preference profile not by looking at voters individually, but by looking at time slots. Note that this does not change the preference profile: it is just another way of looking at the preference profile. For each time slot between $1$ and $n$, each voter $i$  selected a task that she has scheduled in this time slot in her preferred schedule. We call \emph{choice} a triplet $(\mathcal{V}_x,j,t)$ indicating that voter $x$ schedules task $j$ between time $t-1$ and $t$  in her preference $\mathcal{V}_x$. We can thus express a preference profile as a set of choices, such that there are $v$ choices for each time slot and there are $n$ choices for each voter, each task and each slot being chosen exactly once by each voter. We denote by $\mathcal{C}$ the set of all the choices and, for each $y \in \{1 \dots n\}$,  we denote by $\mathcal{C}_y$ the set of all choices $(\mathcal{V}_x,j,t)$ such that $t \leq y$.

\medskip

\noindent {\bf {Iterative breakdown of the tardiness criterion.}}
As we have seen, the total tardiness of a schedule $S$ given a preference profile is the sum, over all voters, of the tardiness of each task in $S$ in comparison to its completion time in the preference of the voter. By breaking down the set of preferences into choices, it is possible to express the tardiness in another way, that will facilitate the analysis of the algorithm. If a task $j$ has been scheduled by a voter $i$ at time slot $t$, then, if it is not scheduled in a solution $S$ by time $t$ then we count a penalty; if it is not scheduled by time $t+1$, then we count another penalty; and so forth. We can then split the tardiness by looking at the tasks scheduled by $S$ at each time slot: for each slot between   $C_j(\mathcal{V}_i)$ to $n$, if task $j$ has not been scheduled yet, then we count $1$ tardiness penalty (for voter $i$). We sum this over all the voters. By this way, we compute for each slot $t$ the number of penalties caused by the decision taken in $t$ -- there will be 1 penalty of each couple (voter $i$, task $j$) if task $i$ has not been completed at time $t$ whereas $C_j(\mathcal{V}_i)\leq t$.


\begin{example}
Let us consider an instance with 5 voters and 5 tasks as follows. Each line represents the preferred schedule of a voter -- e.g. the preferred schedule of the first voter is made of task 1, then task 4, followed by task 2, then task 3 and finally task 5 (such a schedule can be written as: $1\prec 4\prec 2 \prec 3 \prec 5$): 
\begin{figure}[H]
\centering
\begin{tikzpicture}
\task{$1$}{1}{1}{2.4}
\task{$4$}{1}{2}{2.4}
\task{$2$}{1}{3}{2.4}
\task{$3$}{1}{4}{2.4}
\task{$5$}{1}{5}{2.4}

\task{$1$}{1}{1}{1.8}
\task{$5$}{1}{2}{1.8}
\task{$3$}{1}{3}{1.8}
\task{$4$}{1}{4}{1.8}
\task{$2$}{1}{5}{1.8}

\task{$1$}{1}{1}{1.2}
\task{$2$}{1}{2}{1.2}
\task{$3$}{1}{3}{1.2}
\task{$4$}{1}{4}{1.2}
\task{$5$}{1}{5}{1.2}

\task{$2$}{1}{1}{0.6}
\task{$1$}{1}{2}{0.6}
\task{$3$}{1}{3}{0.6}
\task{$5$}{1}{4}{0.6}
\task{$4$}{1}{5}{0.6}

\task{$3$}{1}{1}{0}
\task{$4$}{1}{2}{0}
\task{$1$}{1}{3}{0}
\task{$5$}{1}{4}{0}
\task{$2$}{1}{5}{0}

\draw[->](0-0.5,0)--(5+0.2,0);
\draw (0, 0.1)--(0,-0.1);
\end{tikzpicture}
\end{figure}
Looking at time slot $1$ (between dates $0$ and $1$), task $1$ has been scheduled three times, task $2$ once and task $3$ once. In a solution $S$, scheduling task $1$ at slot $1$ causes a (total) tardiness of $2$ since task $2$ and $3$ which were chosen by two voters will not be scheduled on time. Scheduling task $2$ or task $3$ causes a tardiness of $4$, and scheduling task $4$ or task $5$ creates a tardiness of $5$. 
\label{ex:tardiness_choice}
\end{example}

In the proof of the following proposition,  to compute the sum of the tardiness (also called the total tardiness) of a schedule $S$, we will look a time slots, starting from the first one, between dates 0 and 1, to the last one, between dates $n-1$ and $n$. When looking at time slot $y$, for each choice $(\mathcal{V}_x,j,t) \in \mathcal{C}_y$, we will count a penalty if task $j$ has not been scheduled at time slot $y$ or before. We denote by $k_y$ the number of late tasks at time slot $y$: $k_y(S,P)=\sum_{(\mathcal{V}_x,j,t) \in \mathcal{C}_y} \mathbbm{1}_{C_j(S)>y}$. The total tardiness can be expressed as follows: $T(S,P)=\sum_{y=1}^n k_y(S,P)$.

\begin{prop}
The EMD rule is $2$-approximate for the total tardiness criterion. 
\label{prop:emd_2_approx}
\end{prop}

\begin{proof}
Let us consider a preference profile $P$. Let $S$ be the schedule returned by the EMD rule and let $S^*$ be a schedule minimizing the total tardiness with respect to preference profile $P$. We prove this result by showing that for all $i \geq 0$, $k_i(S,P) \leq 2 k_i(S^*,P)$. 

For $i=0$, we have $k_i(S,P)=k_i(S^*,P)=0$, since no task is scheduled before the first time slot. For any time slot from $1$ to $n$, we express $k_i(S,P)$ as the difference between $i \times v$, the total number of choices from time slot $1$ to time slot $i$, and the number of choices $(\mathcal{V}_x,j,t)$ such that $t \leq i$ and $j$ has been scheduled at the latest at time slot $i$ in $S$.
We denote by $q_i$ the number of tasks with median completion time smaller than or equal to $i$. There are two cases:
\begin{enumerate}
    \item $q_i\leq i$: in this case, the EMD rule schedules  the $q_i$ tasks with median completion time smaller than or equal to $i$ in the $i$ first time slots.  
    Let  $q_i^*$ be the number of tasks with median completion time smaller than or equal to $i$ that are scheduled in $S^*$ at the latest at date $i$. 
    These $q_i^*$ tasks are necessarily scheduled before date $i$ by the EMD rule as well.
    Let $\mathcal{Q}_i^*$ be the set of the $q_i^*$ tasks of median completion time smaller than or equal to $i$ and that are scheduled in $S^*$ before or at time $i$. 
   
    Finally, we denote by $Q_i^*$ the number of choices $(\mathcal{V}_x,j,y)$ of $\mathcal{C}_i$ such that $j \in \mathcal{Q}_i^*$. These choices are removed from the $i \times v$ choices for both the solutions $S$ and $S^*$. 
    There are $q_i-q_i^*$ tasks of median completion time smaller than or equal to $i$ that are scheduled in $S$ before or at time $i$ and that are scheduled after $i$ in $S^*$. For each of these tasks, there are at least $v/2$ choices among the $i\times v$ which are removed by scheduling the task before date $i$.
    There are also $(i-q_i)$ tasks with median completion time strictly larger than $i$ that are scheduled at the latest at date $i$ in $S$, but we have no guarantee that scheduling these tasks remove any choice. We therefore have:  $$k_i(S,P) \leq i\times v-Q_i^*-(q_i-q_i^*)v/2$$
    In $S^*$, there are $(i-q_i^*)$ tasks of median strictly larger than $i$, at most, scheduling these tasks before or at time $i$ removes $v/2$ choices. We then have: $$k_{i}(S^*,P) \geq i\times v-Q_i^*-(i-q_i^*)v/2$$ We then compute: 
    $$
    2k_{i}(S^*,P)-k_{i}(S,P)\geq 2i\times v-2Q_i^*-(i-q_i^*)v-i\times v+Q_i^*+(q_i-q_i^*)v/2
    $$
    $$
    2k_{i}(S^*,P)-k_{i}(S,P)\geq q_i^*v-Q_i^*+(q_i-q_i^*)v/2
    $$
    We know that $Q_i^*\leq q_i^*v$ since each task in $\mathcal{Q}_i^*$ is scheduled at most $v$ times in the preference profile, once per voter. We also know that $q_i \geq q_i^*$. We then have:
    $$
    2k_{i}(S^*,P)-k_{i}(S,P)\geq 0
    $$   
    \item $q_i>i$: in this case, the EMD rule  schedules, from dates $0$ to $i$, exactly $i$ tasks of median completion time smaller than or equal to $i$. There remains $(q_i-i)$ tasks of median completion time smaller than or equal to $i$ that are not scheduled by date $i$ in $S$, the schedule returned by the EMD rule. Each of these tasks can appear in at most $v$ choices in $\mathcal{C}_i$.
    
    Let $r_i\geq 0$ be the number of tasks with median completion time strictly larger than $i$ that are not scheduled by date $i$ in $S$. Let  $\mathcal{R}_i$ the set of these $r_i$ tasks,  and let $R_i$ the set of choices $(\mathcal{V}_x,j,t)$ in $\mathcal{C}_i$ such that $j \in \mathcal{R}_i$. 
    We have: $$k_i(S,P) \leq (q_i-i)v+ |R_i|$$
    In $S^*$, the $i$ tasks scheduled by time slot $i$ are split between the $q_i$ tasks of median completion time smaller than or equal to $i$ and the $r_i$ tasks of median completion time strictly larger than $i$. 
    We denote by $\mathcal{R}_i^*$ the tasks of  $\mathcal{R}_i$ scheduled by $S^*$ before or at time $i$. We call $r_i^*=|\mathcal{R}_i^*|$, and $R_i^*$ the set of choices $(\mathcal{V}_x,j,t)$ of $\mathcal{C}_i$ such that $j\in \mathcal{R}_i^*$. 
    There are $(q_i-(i-r_i^*))$ tasks of median completion time smaller than or equal to $i$ that are not scheduled by $S^*$ by time $i$. Each of these tasks is at least in $v/2$ choices in $\mathcal{C}_i$. We can then write: $$k_i(S^*,P) \geq (q_i-i+r_i^*)v/2+|R_i|-|R_i^*|$$ We then have:
    $$
    2 k_{i}(S^*,P)-k_{i}(S,P) \geq (q_i-i+r_i^*)v +2|R_i|-2|R_i|^*-(q_i-i)v-|R_i|
    $$
    $$
    2 k_{i}(S^*,P)^*-k_{i}(S,P) \geq r_i^*\times v +|R_i|-2|R_i^*|
    $$
    Since $|R_i|\geq |R_i^*|$ and $r_i^*\times v>|R_i^*|$, we  have: $2k_{i}(S^*,P)-k_{i}(S,P) \geq 0$.
\end{enumerate}
In both cases, we have $k_{i}(S,P)\leq 2k_{i}(S^*,P)$ for all $i \geq 1$. Therefore, we have: $\sum_{i=1}^n k_i(S,P) \leq 2 \sum_{i=1}^n k_i(S^*,P)$ and then $\Sigma T(S,P)\leq 2 \Sigma T(S^*,P)$.
\end{proof}

As seen above, since, with unitary tasks, $T(S,P)=  E(S,P)$ and $Dev(S,P)= 2 T(S,P)$, for any schedule $S$ and preference profile $P$, we get the following corollary.   

\begin{corol}
The EMD rule is 2-approximate for the \sigmaD criterion, and for the \sigmaE criterion.
\end{corol}

\paragraph{Additional results in voting theory}
As noted by~\cite{pascual2018collective}, the \sigmaD criterion is equivalent to the Spearman ranking correlation coefficient when task are of unit size, this gives us the following corrolary. Although this minimization problem is polynomially solvable, it is still an interesting property to have for the EMD rule.

\begin{corol}
The EMD rule is 2-approximate for the minimization of the total Spearman correlation coefficient to the preference profile.
\end{corol}




We can show that EMD is 4-approximate for the well-studied Kemeny rule~\cite{kemeny1959mathematics}. The Kemeny rule returns a ranking $R^*$ that minimizes the Kendall-Tau distance $\delta(R,P)$ of a ranking $R$ to the preference profile $P$ where this distance is defined as the number of ordered pairs on which the ranking disagrees with the voters in $P$. 

\begin{prop}
The EMD rule is 4-approximate for the minimization of the Kendall-Tau distance to the preference profile.
\end{prop}

\begin{proof}
~\cite{diaconis1977spearman} showed that for any ranking $R$ and any preference profile $P$, the Spearman correlation coefficient $\rho$ fulfills the following property: $\delta(R,P) \leq \rho(R,P) \leq 2 \delta(R,P)$. 
We call $S$ the solution returned by EMD, $S^*_{KT}$ a solution minimizing the Kendall-Tau distance to the preference profile and $S^*$ a solution minimizing the total Spearman correlation coefficient with the preference profile. 

For the sake of contradiction, let us assume that:

$$
\delta(S,P) > 4 \delta(S^*_{KT},P)
$$

We then have

$$
\rho(S,P) \geq \delta(S,P) > 4 \delta(S^*_{KT},P) \geq 2 \rho(S^*_{KT},P)
$$

And since $S^*$ is optimal for the Spearman, rule we have:
$$
\rho(S,P) \geq \delta(S,P) > 4 \delta(S^*_{KT},P) \geq 2 \rho(S^*_{KT},P) \geq 2\rho(S^*,P)
$$

A contradiction, given the result from Proposition~\ref{prop:emd_2_approx}. We therefore have:
$$
\delta(S,P) \leq 4 \delta(S^*_{KT},P)
$$
\end{proof}

In the next section, we focus on release time and due dates constraints.

\section{Scheduling tasks with time constraints}
\label{sec:timeconstraints}

We first show that it is still possible to compute in polynomial time an optimal solution of the total dissatisfaction of the voters with both the Binary criterion and the Distance criterion presented in Section~\ref{sec:preliminaries}. 

\subsection{Getting optimal solutions with time constraints}

Let us consider that  each task $j\in J$ has a release date $r_j$ and a due date $d_j$. These dates can be  imposed, for example when the tasks represent events that cannot occur before a date $r_j$ or after a date $d_j$. They can also be inferred from the preferences of the voters (by setting $r_j=\min_{i\in\{1,\dots,v\}} \{ r_j(\mathcal{V}_i) \}$  and $d_j=\max_{i\in\{1,\dots,v\}} \{ d_j(\mathcal{V}_i) \}$). In this case, we want no task to be scheduled earlier than in the preferred interval of any voter, or later than in the preferred interval of any voter. This case is particularly interesting if voters are aware of real time constraints on the events that are represented by the tasks, and if the scheduler does not necessarily know these constraints. 

Returning an optimal schedule for both the Binary criterion and the Distance criterion  is, as without any time constraints, an assignment problem. In the bipartite graph with the tasks on the left and the time slots on the right, for each couple (task $j$, time slot $s$), we just put an edge between $j$ and $s$ if and only if $r_j\leq s\leq d_j-1$. The costs of the edges are  equal to the sum of the dissatisfaction of the $v$ voters if task $j$ is scheduled between $s-1$ and $s$. An optimal solution which fulfills time constraints -- if there is one feasible solution -- is a minimum cost matching. Such a matching, if it exists, can be found with Hungarian algorithm in $O(n^3)$~\citep{Tomizawa71,EdmondsK72}.
\medskip

In the next section, we study to which extent the rules presented earlier propagate constraints fulfilled by the preferences of the voters. For example, if all the voters schedule a task after a given time, it may be because this task is not available before. This is particularly interesting in contexts in which the preferences given are not necessarily votes but feasible solutions for a problem (potentially optimizing different aspects like cost, employee satisfaction, inventory management $\dots$). In this case, the question becomes: given several feasible solutions satisfying a set of constraints, do the aggregation rule ensures that the returned solution fulfills the same constraints?
 
\subsection{Axiomatic study of rules with inferred  time constraints}
\label{sec:inferring_constraints}

\subsubsection{Release dates and deadlines consistencies.}

The idea of the two following axioms is the following one: if a task $j$ consistently starts after (resp. ends before) a given date $t$, we can interpret it as $t$ being a firm release date (resp. deadline) for task $j$. In this case, we would like the rule to return a solution in which $j$ starts after (resp. ends before) $t$.

\begin{definition}
Let $V$ be a set of voters and $j$ a task such that for each preference $\mathcal{V}_i$ expressed by voter $i\in\{1, \dots, v\}$, we have $C_j(\mathcal{V}_i) \geq t$, with $t$ a constant. An aggregation rule fulfills \emph{release date consistency} if it always returns a schedule $S$ in which $C_j(S)\geq t$.
\label{def:release_date_consistency}
\end{definition}

\begin{definition}
Let $V$ be a set of voters and $j$ a task such that for each preference $\mathcal{V}_i$ expressed by voter $i\in\{1, \dots, v\}$, we have $C_j(\mathcal{V}_i) \leq t$, with $t$ a constant. An aggregation rule fulfills \emph{deadline consistency} if it always returns a schedule $S$ in which $C_j(S)\leq t$.
\label{def:deadline_consistency}
\end{definition}

We show that the Distance Criterion rule, the Binary Criterion rule, and the EMD rule do not fulfill these axioms. 

\begin{prop}
The \critdistance rule does not fulfill the deadline consistency nor the release date consistency, even when preferences are expressed as schedules.
\label{prop:distance_deadline_release_date_consistency}
\end{prop}

\begin{proof}
Let us consider an instance with 8 tasks and 6 voters and the following preferences:
\begin{figure}[H]
\centering
\begin{tikzpicture}
\task{$6$}{1}{1}{3}
\task{$2$}{1}{2}{3}
\task{$3$}{1}{3}{3}
\task{$4$}{1}{4}{3}
\task{$5$}{1}{5}{3}
\task{$i$}{1}{6}{3}
\task{$j$}{1}{7}{3}
\task{$1$}{1}{8}{3}

\task{$1$}{1}{1}{2.4}
\task{$6$}{1}{2}{2.4}
\task{$3$}{1}{3}{2.4}
\task{$4$}{1}{4}{2.4}
\task{$5$}{1}{5}{2.4}
\task{$i$}{1}{6}{2.4}
\task{$j$}{1}{7}{2.4}
\task{$2$}{1}{8}{2.4}

\task{$1$}{1}{1}{1.8}
\task{$2$}{1}{2}{1.8}
\task{$6$}{1}{3}{1.8}
\task{$4$}{1}{4}{1.8}
\task{$5$}{1}{5}{1.8}
\task{$i$}{1}{6}{1.8}
\task{$j$}{1}{7}{1.8}
\task{$3$}{1}{8}{1.8}

\task{$1$}{1}{1}{1.2}
\task{$2$}{1}{2}{1.2}
\task{$3$}{1}{3}{1.2}
\task{$6$}{1}{4}{1.2}
\task{$5$}{1}{5}{1.2}
\task{$j$}{1}{6}{1.2}
\task{$i$}{1}{7}{1.2}
\task{$4$}{1}{8}{1.2}

\task{$1$}{1}{1}{0.6}
\task{$2$}{1}{2}{0.6}
\task{$3$}{1}{3}{0.6}
\task{$4$}{1}{4}{0.6}
\task{$6$}{1}{5}{0.6}
\task{$j$}{1}{6}{0.6}
\task{$i$}{1}{7}{0.6}
\task{$5$}{1}{8}{0.6}

\task{$1$}{1}{1}{0}
\task{$2$}{1}{2}{0}
\task{$3$}{1}{3}{0}
\task{$4$}{1}{4}{0}
\task{$5$}{1}{5}{0}
\task{$j$}{1}{6}{0}
\task{$i$}{1}{7}{0}
\task{$6$}{1}{8}{0}

\draw[->](0-0.5,0)--(8+0.2,0);
\draw (0, 0.1)--(0,-0.1);


\end{tikzpicture}
\label{fig:profile_couter_example_sigmaD_dates_limite}
\end{figure}
The schedules $(1 \prec 2 \prec 3 \prec 4 \prec 5 \prec 6 \prec i \prec j)$ and $(1 \prec 2 \prec 3 \prec 4 \prec 5 \prec 6 \prec j \prec i)$ are the only two optimal schedules, with a total distance of  $54$. 
They do not fulfill release date consistency since all the voters have completed tasks $i$ and $j$ at time 7 in their preferred schedules, whereas in the returned solution, one of these two tasks is completed at time 8. The best solutions fulfilling release date consistency are $(1 \prec 2 \prec 3 \prec 4 \prec 5 \prec j \prec i \prec 6)$, and $(1 \prec 2 \prec 3 \prec 4 \prec 5 \prec i \prec j \prec 6)$, with a total distance of $56$. 
Therefore, the \critdistance rule does not fulfill deadline consistency.

By reversing the preferences (e.g. , when a preferred schedule $(1 \prec 2 \prec 3 \prec 4 \prec 5 \prec j \prec i \prec 6)$ becomes $(6 \prec i \prec j \prec 5 \prec 4 \prec 3 \prec 2 \prec 1)$), we obtain an instance in which tasks $i$ and $j$ always start after or at $1$, but in which the optimal solutions are  $(i \prec j \prec 6 \prec 5 \prec 4 \prec 3 \prec 2 \prec 1)$ and $(j \prec i \prec 6 \prec 5 \prec 4 \prec 3 \prec 2 \prec 1)$ (the previous optimal solutions but reversed). Either task $i$ or $j$ starts at time 0 in these solutions, whereas no voter schedule theses tasks before time 1. Therefore, the \critdistance rule does not fulfill release date consistency.
\end{proof}

\begin{prop}
The \critbinary  rule does not fulfill the deadline consistency nor the release date consistency. 
\label{prop:deadline_consistencyBin}
\end{prop}

\begin{proof}
Let us consider the following preferences of 3 voters over 7 tasks:
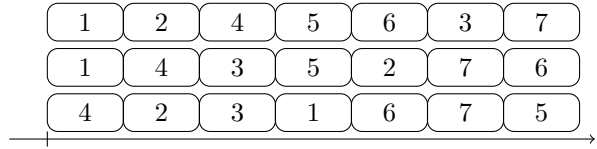
\begin{figure}[H]
\centering
\begin{tikzpicture}

\task{$1$}{1}{1}{1.2}
\task{$2$}{1}{2}{1.2}
\task{$4$}{1}{3}{1.2}
\task{$5$}{1}{4}{1.2}
\task{$6$}{1}{5}{1.2}
\task{$3$}{1}{6}{1.2}
\task{$7$}{1}{7}{1.2}

\task{$1$}{1}{1}{0.6}
\task{$4$}{1}{2}{0.6}
\task{$3$}{1}{3}{0.6}
\task{$5$}{1}{4}{0.6}
\task{$2$}{1}{5}{0.6}
\task{$7$}{1}{6}{0.6}
\task{$6$}{1}{7}{0.6}

\task{$4$}{1}{1}{0}
\task{$2$}{1}{2}{0}
\task{$3$}{1}{3}{0}
\task{$1$}{1}{4}{0}
\task{$6$}{1}{5}{0}
\task{$7$}{1}{6}{0}
\task{$5$}{1}{7}{0}

\draw[->](0-0.5,0)--(7+0.2,0);
\draw (0, 0.1)--(0,-0.1);


\end{tikzpicture}
\label{fig:profile_couter_example_sigmaU_dates_limite}
\end{figure}

We first consider that the deadlines are the one given in the above schedules, and that the release dates are 0. The binary criterion this corresponds to the \sigmaU criterion (which minimizes the number of late tasks). 

The only optimal solution for \sigmaU is $(1 \prec 2 \prec 3 \prec 5 \prec 6 \prec 7 \prec 4)$. The number of late tasks in this solution is 3, task 4 being considered late by the three voters. In a solution fulfilling the deadline consistency property, task 4 has to be completed at most at time 3. This implies that either task 1, task 2 or task 3 has to end after time 3 and will therefore be late for two voters. Additionally, if task 1 is delayed, because of deadline consistency, it has to end at time 4, meaning that task 5 has to be delayed and will therefore be considered late for 2 voters, which amounts to 4 late tasks, more than the optimum. The same line of reasoning can be applied for tasks 2 and 3 if they are delayed after time 3, causing delay to task 5, 6 or 7 if they are scheduled at time 4, 5 or 6. Any solution respecting the deadline consistency property has therefore a number of late tasks of at least 4: the \sigmaU rule does not fulfill deadline consistency.

We can show similarly that the Binary Criterion rule does not fulfill the release date consistency. To this end, we consider that the the deadlines are $n$, and that the release dates are the one given in the above schedules, once they have been reversed. The binary criterion in this case  minimizes the number of early tasks). 
\end{proof}

\begin{prop}
The EMD rule does not fulfill deadline consistency nor release date consistency.
\label{prop:emd_deadline_release_date_consistency}
\end{prop}

\begin{proof}
 Let us consider the following preferences of 3 voters over 4 tasks:

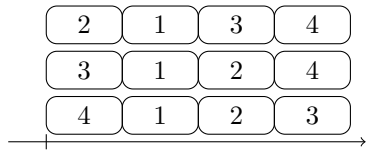
\begin{figure}[H]
\centering
\begin{tikzpicture}

\task{$2$}{1}{1}{1.2}
\task{$1$}{1}{2}{1.2}
\task{$3$}{1}{3}{1.2}
\task{$4$}{1}{4}{1.2}

\task{$3$}{1}{1}{0.6}
\task{$1$}{1}{2}{0.6}
\task{$2$}{1}{3}{0.6}
\task{$4$}{1}{4}{0.6}

\task{$4$}{1}{1}{0}
\task{$1$}{1}{2}{0}
\task{$2$}{1}{3}{0}
\task{$3$}{1}{4}{0}

\draw[->](0-0.5,0)--(4+0.2,0);
\draw (0, 0.1)--(0,-0.1);

\end{tikzpicture}
\label{fig:couter_example_EMD_release_date_consistency}
\end{figure}
With such preferences, the median completion times are as follows: $m_1=2$, $m_2=m_3=3$ et $m_4=4$. The EMD rule returns a schedule in which task $1$ is scheduled first and therefore completes at time $1$, which is before its completion time in all the preferences of the voters. Therefore, the EMD rule does not fulfill release date consistency.

\medskip
 Let us now consider the following preferences of 3 voters over 4 tasks:

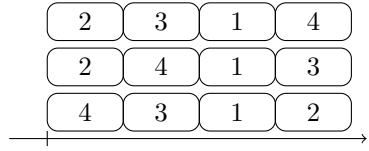
\begin{figure}[H]
\centering
\begin{tikzpicture}

\task{$2$}{1}{1}{1.2}
\task{$3$}{1}{2}{1.2}
\task{$1$}{1}{3}{1.2}
\task{$4$}{1}{4}{1.2}

\task{$2$}{1}{1}{0.6}
\task{$4$}{1}{2}{0.6}
\task{$1$}{1}{3}{0.6}
\task{$3$}{1}{4}{0.6}

\task{$4$}{1}{1}{0}
\task{$3$}{1}{2}{0}
\task{$1$}{1}{3}{0}
\task{$2$}{1}{4}{0}

\draw[->](0-0.5,0)--(4+0.2,0);
\draw (0, 0.1)--(0,-0.1);

\end{tikzpicture}
\label{fig:profile_couter_example_deadlineandreleasedates consistency_EMD}
\end{figure}

With such preferences, the median completion times are as follows: $m_1=3$, $m_2=1$, $m_3=2$ et $m_4=2$. The EMD rule returns a schedule in which task $1$ is scheduled last and therefore completes at time $4$, which is after its completion time in all the preferences of the voters: the EMD rule does not fulfill deadline consistency.
\end{proof}

Since our three rules do no fulfill release date nor deadline consistency, we propose a weaker, yet meaningful, property called \emph{temporal unanimity}.

\subsubsection{Temporal unanimity}

 An aggregation rule satisfies temporal unanimity if, when all voters agree on the time interval during which a task $i$ is scheduled, then $i$ is scheduled during this time interval in the solution returned by the rule. When preferences are given as schedules, this property means that if all voters schedule task $i$ at the same time slot in their preferred schedules, then $i$ should be scheduled at the same time slot in the returned solution. When preferences are expressed as time intervals, it means that if all voters agree on the same release date and deadline for $i$, then $i$ should be scheduled in this given interval in the returned solution. 


\begin{definition}
Let $V$ be a set of voters, and let $j$ be a task such that for each voter $i$ in $V$, we have $d_j(\mathcal{V}_i)=d$, with $d$ a constant, and $r_j(\mathcal{V}_i)=r$, with $r$ a constant strictly smaller than $d$. An aggregation rule fulfills \emph{temporal unanimity} if it returns a schedule in which task $j$ is executed between $r$ and $d$.
\end{definition}

We show that EMD does not fulfill this property, whereas the Binary and the Distance Criterion rules do fulfill this axiom. 

\begin{prop}
The EMD rule does not fulfill the temporal unanimity property.
\label{prop:emd_temporal_unanimity}
\end{prop}

\begin{proof}
Let us consider the following instance:

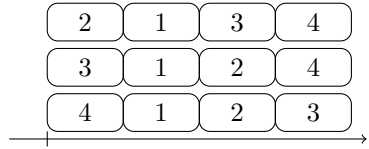
\begin{figure}[H]
\centering
\begin{tikzpicture}

\task{$2$}{1}{1}{1.2}
\task{$1$}{1}{2}{1.2}
\task{$3$}{1}{3}{1.2}
\task{$4$}{1}{4}{1.2}

\task{$3$}{1}{1}{0.6}
\task{$1$}{1}{2}{0.6}
\task{$2$}{1}{3}{0.6}
\task{$4$}{1}{4}{0.6}

\task{$4$}{1}{1}{0}
\task{$1$}{1}{2}{0}
\task{$2$}{1}{3}{0}
\task{$3$}{1}{4}{0}

\draw[->](0-0.5,0)--(4+0.2,0);
\draw (0, 0.1)--(0,-0.1);

\end{tikzpicture}
\label{fig:profile_couter_example_temporal_unanimity_EMD}
\end{figure}
The median completion times are as follows:
$m_1=2$, $m_2=m_3=3$ and $m_4=4$. 
The EMD rule returns a schedule in which task $1$ is scheduled first and thus completes at time $1$ even though it completed at time $2$ in all the schedules expressed by the voters. 
\end{proof}

\begin{prop}
The \critbinary rule fulfills temporal unanimity.
\label{prop:temporal_unanimity_binary}
\end{prop}

\begin{proof}
Let us consider a task ${s(1)}$ such that, for all voter $i$, $d_{s(1)}(\mathcal{V}_i)=d$ and $r_{s(1)}(\mathcal{V}_i)=r$ with $r$ and $d$ two constants such that $0\leq r <d \leq n$. Let us now consider an optimal schedule $S^*$ for the \critbinary minimization in which task ${s(1)}$ is not scheduled between $r$ and $d$. Since each of the preferences has to be compatible with a feasible schedule, in each preferences there are at most $d-r$ tasks with release dates and deadlines included in the $[r,d]$ interval. Since task ${s(1)}$ is always included in this interval in the preferences of the voters, there is in the $[r,d]$ interval of $S^*$ at least one task $s(2)$ is scheduled in the preferences of the voters at least once before $r$ or after $d$. We distinguish two sub-cases. 
\begin{itemize}
    \item If this task $s(2)$ does not have a unique release date $r'$ given by the voters and a unique due date $d'$ given by the voters, then we can simply perform the swap between the positions of ${s(1)}$ and ${s(2)}$ to obtain a new solution $S'$ in which there is one less task out of its unique time interval and which is at least as good as $S^*$ since we decrease the binary criterion cost for ${s(1)}$ by the number of voters $v$ and we increase it for ${s(2)}$ by at most $v$.
    \item If this task $s(2)$ has a unique release date $r'$ and due date $d'$ then we consider two subcases:
    \begin{itemize}
        \item If task ${s(2)}$ is scheduled in $S^*$ before $r'$ or after $d'$, and therefore not in its unique time interval,  or if its time interval covers the position of ${s(1)}$ in $S^*$, we can perform the swap between the positions of ${s(1)}$ and ${s(2)}$ as in the above mentioned  case.
        \item Otherwise, we consider the other tasks, if any, scheduled in $S^*$ between $r$ and $d$ and which are not always scheduled between $r$ and $d$ in the preferences. If none of these tasks fulfill any of the two previous conditions then we consider the set $\mathcal{T}_{s(1)}$ of all these tasks scheduled between $r$ and $d$ in $S^*$ and which have a unique time interval in which they are scheduled. For each of these tasks, its time interval $r',d'$ is either as $r'<r$ or $d'>d$, or both. We now consider the interval from the smallest unique release date of a task in $\mathcal{T}_{s(1)}$ to the maximum unique deadline of a task in $\mathcal{T}_{s(1)}$. We then repeat the same reasoning as above: 
        \begin{itemize}
            \item if there is a task ${s(3)}$ which is in the time interval of a task ${s(2)}$ from $\mathcal{T}_{s(1)}$, and which does not have a unique time interval or which has a time interval containing the position of ${s(1)}$ in $S^*$, then we perform the following circular exchange: task ${s(1)}$ takes the time slot of ${s(2)}$, which takes the time slot of ${s(3)}$, which takes the time slot of ${s(1)}$. Such a circular exchange does not increase the binary criterion, since the cost relative to ${s(1)}$ is decreased by $v$, the cost relative to ${s(2)}$ does not increase, since ${s(2)}$ stays in its interval, and the cost of ${s(3)}$ increases by at most $v$.
            \item  If no such task  ${s(3)}$ exists, then there is at least one task which has a unique release date $r''<r'$, or a unique deadline $d''>d'$, or both. We then consider the set $\mathcal{T}_{s(2)}$ of such tasks and expand the considered interval. Since at each of these steps we extend the considered interval by at least one unit of time, the interval will necessarily include the position of ${s(1)}$ at some point and we will be able to perform a swap.
        \end{itemize}
            \end{itemize}
\end{itemize}
\end{proof}

\begin{prop}
The \critdistance rule fulfills temporal unanimity when preferences are schedules.
\label{prop:temporal_unanimity_distance_shcedule}
\end{prop}

\begin{proof}
Let us consider that a task $l$ is always scheduled between time $k$ and time $k+1$ in the preferences of the voters. Let $S^*$ be an optimal solution for the \critdistance minimization, and let us assume, by contradiction, that task $l$ is not scheduled between $k$ and $k+1$ in $S^*$. Let $S$ be a schedule obtained from $S^*$ by swapping the positions of task $l$ and the task $j$ scheduled between $k$ and $k+1$ in $S^*$. Note that the distance of any task other than $l$ or $j$ is the same in $S$ and $S^*$. The distance of task $l$ is decreased by the absolute value of the difference between its position in $S$ and its position in $S^*$, times the number of voters (since all voters scheduled it between $k$ and $k+1$), while the distance of task $j$ is increased by at most this value. Therefore $S$ is an optimal schedule as well.

    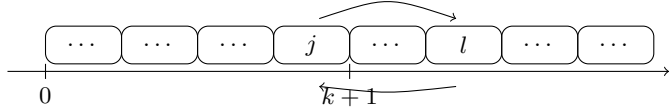
\begin{figure}[H]
    \centering
    \small
    \begin{tikzpicture}
        \task{$\dots$}{1}{1}{0}
        \task{$\dots$}{1}{2}{0}
        \task{$\dots$}{1}{3}{0}
        \task{$j$}{1}{4}{0}
        \task{$\dots$}{1}{5}{0}
        \task{$l$}{1}{6}{0}
        \task{$\dots$}{1}{7}{0}
        \task{$\dots$}{1}{8}{0}

        \draw[->](0-0.5,0)--(8+0.2,0);
        \draw (0, 0.1)--(0,-0.1) node[below]{$0$};
        \draw (4, 0.1)--(4,-0.1) node[below]{$k+1$};

        \draw[->]  (3.6,0.7) .. controls (4.5,1) .. (5.4,0.7);
        \draw[->]  (5.4,-0.2) .. controls (4.5,-0.3) .. (3.6,-0.2);

    \end{tikzpicture}
    \caption{Schedule $S^*$ and the swap performed to obtain $S$.}
    \label{fig:distance_temporal_una_1}
    \end{figure}

Let us now examine the case in which the distance of task $j$ has increased by $|C_{l}(S^*)-C_{j}(S^*)|$ -- we will actually show that this cannot happen. Note that if task $j$ was scheduled before task $l$ in $S^*$ (case 1) then the distance of $j$ increased by $C_{l}(S^*)-C_{j}(S^*)$: it means that $j$ has been scheduled before its completion time in $S^*$ by all voters. Likewise, if task $j$ was scheduled  after  task $l$ in $S^*$ (case 2) then the distance of $j$ increased by  $C_{j}(S^*)-C_{l}(S^*)$: it means that $j$ has been scheduled after its completion time in $S^*$ by all the voters. 


In case 1, let $b$ be the maximum completion time of task $j$ in the preference profile, and let $k$ be the task which is completed at time $b$ in $S^*$. We build schedule $S'$ from $S$ by swapping the position of task $j$ and task $k$. The distance of $j$ is decreased by the difference between the position of $j$ and $k$ for all voters. If the distance of $k$ increases by the same value it means that task $k$ always completes before $b$ in the preferences of the voters. By repeating such swaps, the date $b$ is decreased each time and we will necessarily reach a point where we either find a task for which the distance increase is smaller than the distance decrease when doing the swap or find a $b$ of $1$. 

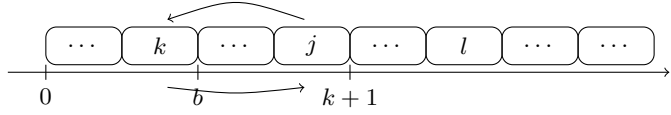
\begin{figure}[H]
    \centering
    \small
    \begin{tikzpicture}
        \task{$\dots$}{1}{1}{0}
        \task{$k$}{1}{2}{0}
        \task{$\dots$}{1}{3}{0}
        \task{$j$}{1}{4}{0}
        \task{$\dots$}{1}{5}{0}
        \task{$l$}{1}{6}{0}
        \task{$\dots$}{1}{7}{0}
        \task{$\dots$}{1}{8}{0}

        \draw[->](0-0.5,0)--(8+0.2,0);
        \draw (0, 0.1)--(0,-0.1) node[below]{$0$};
        \draw (2, 0.1)--(2,-0.1) node[below]{$b$};
        \draw (4, 0.1)--(4,-0.1) node[below]{$k+1$};

        \draw[->]  (3.4,0.7) .. controls (2.5,1) .. (1.6,0.7);
        \draw[->]  (1.6,-0.2) .. controls (2.5,-0.3) .. (3.4,-0.2);

    \end{tikzpicture}
    \caption{Schedule $S^*$ and a preliminary swap (case 1) ensuring that the final swap of task $l$ will strictly decrease the total distance.}
    \label{fig:distance_temporal_una_2}
    \end{figure}

The same thing can be done in case 2, by defining $b$ as the minimum completion time of $j$ in the preference profile, and  $k$ as the task which is completed at time $b$ in $S^*$. The distance of $j$ is decreased by the difference between the position of $j$ and $k$ for all voters. If the distance of $k$ increases by the same value than the distance of $j$ is decreased, it means that task $k$ always completes after $b$ in the preferences of the voters. By repeating such swaps, the date $b$ is increased each time and we will necessarily reach a point where we either find a task for which the distance increase is smaller than the distance decrease when doing the swap or find a $b$ of $n$.

If we did not find $b=1$, or $b=n$, it means that we have found a schedule of cost (sum of the distances) better than $S^*$, a contradiction. If we ended with $b=1$ or $b=n$, then a task that would always complete before (resp. after) or at time $b=1$ (resp. $b=n$) would always be scheduled first (resp. last),  and doing the swap will always be strictly better. The solution obtained after doing the swaps is strictly better than the solution $S^*$ supposed to be optimal, a contradiction. 

\end{proof}

\begin{prop}
The \critdistance rule does not fulfill temporal unanimity when preferences are expressed as release dates and deadlines.
\label{prop:temporal_unanimity_distance_interval}
\end{prop}

\begin{proof}
Let us consider an instance with 8 tasks and 6 voters and the following preferences:
\begin{figure}[H]
\centering
\begin{tikzpicture}
\task{$6$}{1}{1}{3}
\task{$2$}{1}{2}{3}
\task{$3$}{1}{3}{3}
\task{$4$}{1}{4}{3}
\task{$5$}{1}{5}{3}
\task{$i/j$}{2}{7}{3}
\task{$1$}{1}{8}{3}

\task{$1$}{1}{1}{2.4}
\task{$6$}{1}{2}{2.4}
\task{$3$}{1}{3}{2.4}
\task{$4$}{1}{4}{2.4}
\task{$5$}{1}{5}{2.4}
\task{$i/j$}{2}{7}{2.4}
\task{$2$}{1}{8}{2.4}

\task{$1$}{1}{1}{1.8}
\task{$2$}{1}{2}{1.8}
\task{$6$}{1}{3}{1.8}
\task{$4$}{1}{4}{1.8}
\task{$5$}{1}{5}{1.8}
\task{$i/j$}{2}{7}{1.8}
\task{$3$}{1}{8}{1.8}

\task{$1$}{1}{1}{1.2}
\task{$2$}{1}{2}{1.2}
\task{$3$}{1}{3}{1.2}
\task{$6$}{1}{4}{1.2}
\task{$5$}{1}{5}{1.2}
\task{$i/j$}{2}{7}{1.2}
\task{$4$}{1}{8}{1.2}

\task{$1$}{1}{1}{0.6}
\task{$2$}{1}{2}{0.6}
\task{$3$}{1}{3}{0.6}
\task{$4$}{1}{4}{0.6}
\task{$6$}{1}{5}{0.6}
\task{$i/j$}{2}{7}{0.6}
\task{$5$}{1}{8}{0.6}

\task{$1$}{1}{1}{0}
\task{$2$}{1}{2}{0}
\task{$3$}{1}{3}{0}
\task{$4$}{1}{4}{0}
\task{$5$}{1}{5}{0}
\task{$i/j$}{2}{7}{0}
\task{$6$}{1}{8}{0}

\draw[->](0-0.5,0)--(8+0.2,0);
\draw (0, 0.1)--(0,-0.1);


\end{tikzpicture}
\label{fig:profile_couter_example_sigmaD_unanimity}
\end{figure}
In this profile each voter gives a time interval of $1$ for all tasks except for $i$ and $j$ which have a time interval of $2$. For example the first voter indicates that task $6$ has a release date of $0$ and a due date of $1$, while both tasks $i$ and $j$ have a release date of $5$ and a due date of $7$. The two optimal solutions for the \critdistance minimization are $(1 \prec 2 \prec 3 \prec 4 \prec 5 \prec 6 \prec i \prec j)$ and $(1 \prec 2 \prec 3 \prec 4 \prec 5 \prec 6 \prec j \prec i)$, with a total distance of $48$.  
These optimal schedule do not fulfill temporal unanimity since either $i$ or $j$ is scheduled outside of the time interval agreed on by all the voters. The best solutions fulfilling temporal unanimity are $(1 \prec 2 \prec 3 \prec 4 \prec 5 \prec j \prec i \prec 6)$, and $(1 \prec 2 \prec 3 \prec 4 \prec 5 \prec i \prec j \prec 6)$, with a total distance of $50$. 
\end{proof}

\section{Precedence constraints}
\label{sec:precedence}

In this section, we focus on precedence constraints between the tasks. We will consider two settings.

Firstly, we consider a setting in which the precedence constraints are known by the voters. In this setting, that we call \emph{\precedenceinduite}, if a task $a$ has to be scheduled before a task $b$, then, in  the preference $\mathcal{V}_x$ of any voter $x$, we have $C_a(\mathcal{V}_x)<C_b(\mathcal{V}_x)$. Our aim is to determine whether or not a given aggregation rule guarantees that task $a$ will be scheduled before task $b$ in the schedule returned by the rule. Note that in voting theory this property is called \emph{unanimity}. 

The second setting corresponds to the case in which the precedence constraints are not known by the voters, and therefore preferences do not necessarily fulfill these precedence constraints: the precedence constraints only exist for the schedule that has to be returned.  
This setting is called \emph{\precedencegraphe}. 
\medskip

We  define a family of optimization problems of form \textsc{$\alpha-Prec$} where $\alpha$ is an optimization criterion and $Prec$ is a setting for the precedence constraints: it is \textsc{Inferred} when precedence constraints are fulfilled by the preferences, or \textsc{Graph} when the preference constraints only apply to the  returned solution. 
For example, the problem \sigmaDgraph has the following input: a set $J$ of $n$ tasks ; an acyclic directed graph $G$ which represents the precedence constraints between the tasks in $J$ ; a set $V$ of $v$ preferred schedules (permutation of tasks)  -- these schedules do not necessarily fulfill the precedence constraints. The aim is to output a schedule which fulfills the precedence constraints and, among these feasible schedules, which  minimizes the sum of the deviations with respect to the preferences of the voters:
$\sum_{i \in V} \sum_{j \in J} D_j(S,\mathcal{V}_i)$.

\medskip

Our aim is to study the complexity of problems mentioned before (total deviation, total tardiness, number of late tasks), when there are inferred or given precedence constraints. In Section~\ref{subsec:inferred}, we study the case in which precedence constraints are inferred by the preferences of the voters, and we show that problems \sigmaDinf and \sigmaTinf can be solved in polynomial time. In Section~\ref{subsec:notinferred}, we study the case in which precedence constraints are given and are not necessarily fulfilled by the preferred schedules of the voters. We will show that problems \sigmaDgraph, \sigmaTgraph and \sigmaUgraph are NP-hard.

\subsection{Inferred precedence constraints}
\label{subsec:inferred}

\begin{prop}
Problems \sigmaDinf and \sigmaTinf can be solved in $O(vn^2+n^3)$. 
\label{prop:permutations}
\end{prop}

\begin{proof}
~\cite{durand2022collective} showed that when two tasks $a$ and $b$ are of same length, if task $a$ is scheduled before $b$ in all the preferences then there exists an optimal solution for the \sigmaD rule and the \sigmaT rule such that $a$ is scheduled before $b$. Additionally, for any optimal solution in which $b$ would be scheduled before $a$, it is possible to swap the position of $a$ and $b$ without increasing the deviation (or the tardiness). Therefore by doing successive permutations from an optimal solution, we can find another optimal solution in which precedence constraints are fulfilled. We now show that the number of permutations needed is bounded by $n^2$. Problems \sigmaDinf and \sigmaTinf can thus be solved in polynomial time by 
\begin{enumerate}
    \item computing an optimal solution of \sigmaD or \sigmaT (without the precedence constraints) through an assignment problem, as seen in the introduction. This can been done in $O(vn^2+n^3)$; 
    \item swapping couple of tasks $(a,b)$ that do not fulfill precedence constraints in the returned schedule of Step 1. As we will show now, there will be at most $n^2$ swaps. 
\end{enumerate}

We create a precedence directed graph with $n$ vertices, one for each task, and in which there is an edge from  vertex $a$ to  vertex $b$ if the task corresponding to  vertex $a$ is always scheduled before the task corresponding to vertex $b$ in the preferences of the voters. There are at most $n^2$ edges so it is possible to create this graph in $O(n^2)$ operations.
Note that this precedence relation is transitive: if $a$ is always scheduled before $b$ and $b$ is always scheduled before $c$ then $a$ is always scheduled before $c$. This implies that this graph has no cycle. This also implies that there exists at least one vertex with no successor.

We choose a vertex $x$ among the vertices with no successor in the above mentioned precedence graph.  For readability, we will in the sequel denote the task corresponding to vertex $x$ as task $x$. We look whether among the predecessors of $x$ there exist vertices corresponding to a task scheduled after $x$ in the optimal schedule returned by \sigmaD or \sigmaT. If such vertices exist, we swap the position of the task $x$ with the task corresponding to its predecessor scheduled after it and as close as possible to  $x$ in the returned schedule. We repeat this step until all the tasks corresponding to predecessors of $x$ are scheduled before $x$. 
By swapping $x$ with its closest predecessor scheduled after it, we make sure that we do not create any violation of the precedence constraints. 
Studying all the vertices takes $n$ operations, consisting in at most $n$ swaps:  the total number of swaps is then bounded by $n^2$. 
\end{proof}


Note that the previous proof is a constructive proof: we can compute   an optimal solution for \sigmaDinf (or \sigmaTinf) by  solving an assignment problem for \sigmaD (or \sigmaT), and then swapping tasks which do not fulfill the precedence constraints as explained in the proof of Proposition~\ref{prop:permutations}.



We cannot take the same approach for \sigmaUinf. Indeed,  there are instances in which no optimal solution for the minimization of the total number of late tasks criterion fulfills the inferred precedence constraints, as shown by the following proposition.


\begin{prop}
There exist instances for which  no optimal solution for the \sigmaU criterion fulfills the inferred precedence constraints.
\label{prop:sigmaU_precedences_induites}
\end{prop} 

\begin{proof}
Let us consider the following instance of 5 tasks and 6 voters. The number at the left of each schedule indicates the number of voters whose schedule is the preferred schedule (e.g. the favorite schedule of three voters is the second schedule).  
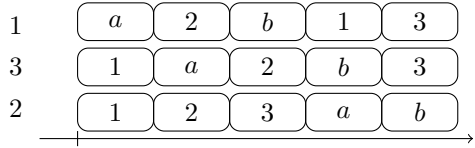
\begin{figure}[H]
\centering
\begin{tikzpicture}

\task{$a$}{1}{1}{1.2}
\task{$2$}{1}{2}{1.2}
\task{$b$}{1}{3}{1.2}
\task{$1$}{1}{4}{1.2}
\task{$3$}{1}{5}{1.2}

\task{$1$}{1}{1}{0.6}
\task{$a$}{1}{2}{0.6}
\task{$2$}{1}{3}{0.6}
\task{$b$}{1}{4}{0.6}
\task{$3$}{1}{5}{0.6}

\task{$1$}{1}{1}{0}
\task{$2$}{1}{2}{0}
\task{$3$}{1}{3}{0}
\task{$a$}{1}{4}{0}
\task{$b$}{1}{5}{0}

\draw[->](0-0.5,0)--(5+0.2,0);
\draw (0, 0.1)--(0,-0.1);

\node[text width=1cm] at (-0.4,1.50) {$1$};
\node[text width=1cm] at (-0.4,0.95) {$3$};
\node[text width=1cm] at (-0.4,0.40) {$2$};
\end{tikzpicture}
\label{fig:profile_couter_example_unanimity_sigmaU}
\end{figure}
The only optimal solution for the \sigmaU criterion is the following one:

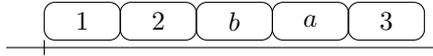
\begin{figure}[H]
\centering
\begin{tikzpicture}
\task{$1$}{1}{1}{0}
\task{$2$}{1}{2}{0}
\task{$b$}{1}{3}{0}
\task{$a$}{1}{4}{0}
\task{$3$}{1}{5}{0}

\draw[->](0-0.5,0)--(5+0.2,0);
\draw (0, 0.1)--(0,-0.1);
\end{tikzpicture}
\label{fig:solution_couter_example_unanimity_sigmaU}
\end{figure}
In this solution, $b$ is scheduled before $a$, whereas all the voters have scheduled $a$ before $b$ in their favorite schedules: this violates the inferred precedence constraints.
\end{proof}

This last proposition means that we cannot proceed like in Proposition~\ref{prop:permutations}, by computing an  optimal solution for \sigmaU and then swapping tasks which would not be in the right order.  Whether  problem \sigmaUinf is NP-hard or not is an open question.


We end this section by noting that the rule EMD computes  schedules which fulfill all the inferred precedence constraints. Indeed, if all the voters schedule a task $a$ before a task $b$ in their preferred schedules, then the median completion time of $a$ will be smaller than the median completion time of $b$ and thus $a$ will be scheduled before $b$ in the schedule returned by EMD. 

\subsection{Imposed precedence graph}
\label{subsec:notinferred}


We start by proving that this problem is strongly NP-hard for the total tardiness criterion.

\begin{prop}
The \sigmaTgraph problem is strongly NP-hard, even when the precedence graph consists in chains.
\label{prop:sigmaT_graph}
\end{prop}

We prove this proposition by doing a polynomial time reduction from the scheduling problem denoted by  \chainsSigmaT using the Graham's notation, a classical way to denote problems in scheduling theory~\citep{brucker1999scheduling}. An instance of this problem is: 
\begin{itemize}
    \item a set $J$ of $n$ unit tasks, each task $j$ having a due date $d_j$. Without loss of generality, we assume that $d_j \leq n$ for all $j$. 
    \item a precedence graph,  modeling precedence constraints between the tasks. We assume that  and this graph is made of chains in chains (i.e. each task has at most one successor and one predecessor, and there are no cycle). 
\end{itemize} 
The optimization version of this problem consists in minimizing the sum of the tardiness of the tasks. The decision version of this problem consists in answering the following question: given an integer $K$, is there a schedule $S$ of the tasks in $J$ on a single machine, such that the precedence constraints are fulfilled, and such that the total tardiness of the tasks, $\sum_{j \in J} \max(0,C_j(S)-d_j)$,  is smaller than  or equal to $K$ ? 
%
This problem is known to be NP-hard~\citep{leung1990minimizing}.

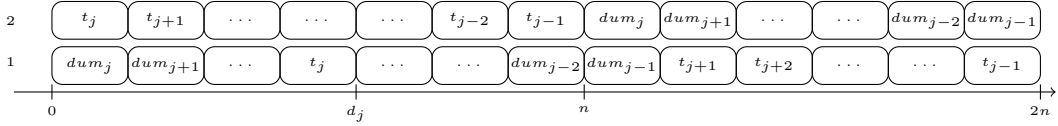
\begin{figure}[H]
\centering
\tiny
\begin{tikzpicture}

\task{$t_j$}{1}{1}{0.6}
\task{$t_{j+1}$}{1}{2}{0.6}
\task{$\dots$}{1}{3}{0.6}
\task{$\dots$}{1}{4}{0.6}
\task{$\dots$}{1}{5}{0.6}
\task{$t_{j-2}$}{1}{6}{0.6}
\task{$t_{j-1}$}{1}{7}{0.6}
\task{$dum_{j}$}{1}{8}{0.6}
\task{$dum_{j+1}$}{1}{9}{0.6}
\task{$\dots$}{1}{10}{0.6}
\task{$\dots$}{1}{11}{0.6}
\task{$dum_{j-2}$}{1}{12}{0.6}
\task{$dum_{j-1}$}{1}{13}{0.6}

\task{$dum_j$}{1}{1}{0}
\task{$dum_{j+1}$}{1}{2}{0}
\task{$\dots$}{1}{3}{0}
\task{$t_j$}{1}{4}{0}
\task{$\dots$}{1}{5}{0}
\task{$\dots$}{1}{6}{0}
\task{$dum_{j-2}$}{1}{7}{0}
\task{$dum_{j-1}$}{1}{8}{0}
\task{$t_{j+1}$}{1}{9}{0}
\task{$t_{j+2}$}{1}{10}{0}
\task{$\dots$}{1}{11}{0}
\task{$\dots$}{1}{12}{0}
\task{$t_{j-1}$}{1}{13}{0}

\draw[->](0-0.5,0)--(13+0.2,0);
\draw (0, 0.1)--(0,-0.1) node[below]{$0$};
\draw (4, 0.1)--(4,-0.1) node[below]{$d_j$};
\draw (7, 0.1)--(7,-0.1) node[below]{$n$};
\draw (13, 0.1)--(13,-0.1) node[below]{$2n$};

\node[text width=1cm] at (-0.1,0.95) {$2$};
\node[text width=1cm] at (-0.1,0.40) {$1$};
\end{tikzpicture}
\caption{Preferred schedules of the $3$ voters generated for task $j$. }
\label{fig:voters_sigmaT}
\end{figure}

We create an instance of \sigmaDgraph from the instance from \chainsSigmaT as follows.
\begin{itemize}
    \item For each task $j$ in $J$ we create a task $t_j$ and a task $dum_j$. These tasks are split into two sets $J_t=\{j_1,\dots, j_n\}$ and $J_{dum}=\{dum_1,\dots, dum_n\}$. The set $J'$ of the tasks of the instance of \sigmaTgraph  is the union of $J_t$ and $J_{dum}$.
    \item  For each precedence relation in the \chainsSigmaT problem between tasks $i$ and $j$, we create a precedence constraint between $t_i$ and $t_j$ in the precedence graph of problem \sigmaTgraph.
    \item 
  For each task $j$ in $J$ we also create three voters. 
Their preferred schedules, that we will describe now, are represented on Figure~\ref{fig:voters_sigmaT}. 
The first two voters, of type $T$, schedule $t_j$ first, followed by $t_{j+1}$ and so forth until $t_n$, then $t_1$ to $t_{j-1}$ by increasing index. They then schedule the $dum$  tasks following the same pattern:  $dum_j$ first, then $dum_{j+1}$ to $dum_n$, followed by $dum_1$ to $dum_{j-1}$ by increasing index (see top schedule in Figure~\ref{fig:voters_sigmaT}). The last voter, of type $D$, schedules task $t_j$ between time $d_j-1$ and $d_j$. Before that,  she schedules $(d_j-1)$ $dum$ tasks from $dum_j$ by increasing index (using again a circular order of the tasks, where task $dum_1$ follows task $dum_n$). The remaining  $dum$ tasks are scheduled after $t_j$ by increasing indexes. The schedule is completed  with tasks $t_{j+1}$, $t_{j+2}$, \dots, until task $t_{j-1}$ if $j\neq 1$, or task $t_n$ if $j=1$  (see bottom schedule in Figure~\ref{fig:voters_sigmaT}). 
\end{itemize} 


In order to prove Proposition~\ref{prop:sigmaT_graph}, we start by proving the following lemma. 

\begin{lemma}
For the instance of the \sigmaTgraph problem described above, there is an optimal solution in which all the $t$ tasks are scheduled before all the $dum$ tasks.
\label{lem:reduction_sigmaT}
\end{lemma}

\begin{proof}
Let us assume by contradiction that there is no optimal solution in which all $t$ tasks are scheduled before all $dum$ tasks. Let $S$ be such an optimal solution: there is at least one $dum$ task completing just before a task $t$. Let us call $dum_i$ the first $dum$ task scheduled before a $t$ task, and $t_j$ be the $t$ task scheduled just after $dum_i$. Let $k$ be the completion time of $dum_i$ in $S$: we have $C_{dum_i}(S)=k$ and $C_{t_j}(S)=k+1$ (note that $1\leq k <2n$).

We call $S'$ the schedule obtained from $S$ by swapping the position of $dum_i$ and $t_j$. The total tardiness of $S'$ is similar to $S$ except for the tardiness of $dum_i$ and $t_j$. We then have $C_{dum_i}(S')=k+1$ and $C_{t_j}(S')=k$. Note that if the precedence constraints over the $t$ tasks are satisfied by $S$, they are also satisfied by $S'$ since the order on the $t$ tasks has not changed. Therefore, since $S$ is a feasible solution, $S'$ is also a feasible solution. We distinguish two sub-cases:

\begin{figure}[H]
\centering
\small
\begin{tikzpicture}

\task{$t_k$}{1}{1}{0.6}
\task{$\dots$}{1}{2}{0.6}
\task{$t_l$}{1}{3}{0.6}
\task{$dum_a$}{1}{4}{0.6}
\task{$\dots$}{1}{5}{0.6}
\task{$dum_i$}{1}{6}{0.6}
\task{$t_j$}{1}{7}{0.6}
\task{$\dots$}{1}{8}{0.6}
\task{$\dots$}{1}{9}{0.6}
\task{$\dots$}{1}{10}{0.6}

\task{$t_k$}{1}{1}{0}
\task{$\dots$}{1}{2}{0}
\task{$t_l$}{1}{3}{0}
\task{$dum_a$}{1}{4}{0}
\task{$\dots$}{1}{5}{0}
\task{$t_j$}{1}{6}{0}
\task{$dum_i$}{1}{7}{0}
\task{$\dots$}{1}{8}{0}
\task{$\dots$}{1}{9}{0}
\task{$\dots$}{1}{10}{0}

\draw[->](0-0.5,0)--(10+0.2,0);
\draw (0, 0.1)--(0,-0.1) node[below]{$0$};
\draw (6, 0.1)--(6,-0.1) node[below]{$k$};

\node[text width=1cm] at (-0.1,0.95) {$S$};
\node[text width=1cm] at (-0.1,0.40) {$S'$};
\end{tikzpicture}
\caption{Schedules $S$ and $S'$. The first $dum$ task to be scheduled just before a $t$ task in $S$ is $dum_i$.}
\label{fig:voters_S_Sprime}
\end{figure}
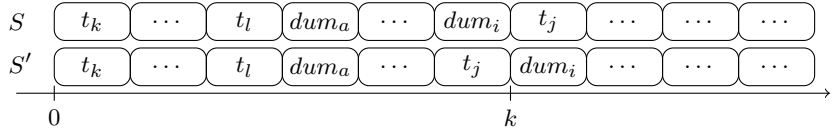

\begin{itemize}
    \item $k \leq n$. In this case, the tardiness relative to task $t_j$ is reduced in $S'$ in comparison to $S$ by at least $2k$. Indeed, there are $2k$ voters scheduling $t_j$ at time $k$ or before. Therefore, moving $t_j$ from $k+1$ to $k$ reduces the tardiness of $t_j$ by one for each of these voters, giving a total of $2k$. 
    The tardiness of task $dum_i$ is increased in $S'$ in comparison to $S$. There are at most $k$ voters of type $D$ scheduling $dum_i$ at time $k$ or before:  for each of these voters, the tardiness is increased by one. 
    The sum of the tardiness of $S'$ is   decreased by at least $2k$ (due to  $t_j$), and increased by at most $k$ (due to  $dum_i$),  in comparison to the sum of the tardiness of $S$:  the total tardiness of $S'$ is thus smaller than the tardiness of $S$. Since  $S$ minimizes the sum of the tardiness, there is a contradiction. 
    \item $k>n$. In this case, the tardiness relative to task $t_j$ is reduced in $S'$ in comparison to $S$ by $2n+1+(k-n)$.  
    Indeed, there are $2n$ voters of type $T$ scheduling $t_j$ before $k$, and one voter of type $D$ scheduling $t_j$ so that this task is completed at date $d_j$ with $d_j \leq n$. This makes a total of $2n+1$. Additionally, there are 
    $k-n$ voters of type $D$ scheduling $t_j$ at time $n+1$, $n+2$ up to $k$. For each of these voters, the tardiness of $t_j$ is reduced by one. 
    
    On the other hand, the tardiness of $dum_i$ is increased in $S'$ in comparison to $S$ by $n+2(k-n)$. The $n$  voters of type $D$ scheduled $dum_i$ so it  that it is completed at most at time $n+1$, meaning that delaying $dum_i$ from $k$ to $k+1$ increases the tardiness by one for each of these $n$ voters. Additionally, there are $2(k-n)$ voters of type $T$ scheduling $dum_i$ so that it completes at dates $n+1$, $n+2$  to $k$: the tardiness is increased by one for each of these voters. If we compare the increase in tardiness for task $dum_i$, $n+2(k-n)$, to the decrease of the tardiness for task $t_j$, $2n+1+(k-n)$, we see that the sum of the tardiness in $S'$ is decreased by $2n+1-k$. Since $k<2n$, this value is always strictly positive. This means that the total tardiness of $S'$ is strictly smaller  than the tardiness of $S$, an optimal solution: a contradiction. 
   \end{itemize}
\end{proof}


We can now start the proof of Proposition~\ref{prop:sigmaT_graph}.

\begin{proof}
From Lemma~\ref{lem:reduction_sigmaT}, we know that there exists an optimal schedule $S$ in which $t$ tasks  are scheduled before $dum$ tasks. We analyze the sum of the tardiness in such a schedule. We first show that the sum of the tardiness of $dum$ tasks is the same in any schedule fulfilling the property of Lemma~\ref{lem:reduction_sigmaT} (first item below), and we then analyze the sum of the tardiness due to $t$ tasks  (second item below). 
\begin{itemize}
    \item We show that in any schedule in which $dum$ tasks are scheduled after $t$ tasks  fulfilling this property, the tardiness due to $dum$ tasks is always the same.
    
    Voters of type $T$ schedule each $dum$ task twice between $n$ and $n+1$, twice between $n+1$ and $n+2$ and so on until $2n-1$ and $2n$. In schedule $S$, the $dum$ task scheduled between $n$ and $n+1$ in not late for any voter of type $T$, the task scheduled between $n+1$ and $n+2$ is late of one unit of time for 2 voters of type $T$, and so on. Overall, the total tardiness of $dum$ tasks for $T$ voters is then $2\sum_{i=1}^n \sum_{j=1}^{i} (j-1)$, a constant number.

    Let us now show that the sum of the tardiness of $dum$ tasks for $D$ voters will be the same in any schedule $S$ in which $t$ tasks  are scheduled before $dum$ tasks. Indeed, for each $D$ voter $j$, and for each task $dum_i$, the completion time of $dum_i$ in the preferred schedule of $j$ is at most $n+1$, whereas the completion time of $dum_i$ in $S$ is at least $n+1$. Therefore, the sum of the tardiness due to $dum$ tasks for $D$ voters is equal to the sum of the distances between completion times of $dum$ tasks in the preferred schedules of voters $D$ to date $n+1$ -- which is a constant, since preferred schedules are fixed --, plus the sum of the distances of $dum$ tasks between date $n+1$ and the completion time of $dum$ tasks in $S$ -- this is also a constant since the completion times of $dum$ tasks in $S$ are the set of times $\{n+1, \dots, 2n\}$. Therefore, the the sum of the tardiness of $dum$ tasks for $D$ voters is a constant.   

    We have seen that the sum of the tardiness of $dum$ tasks is value is the same for any schedule $S$ which fulfills Lemma~\ref{lem:reduction_sigmaT}. Let $T_{dum}$ denote this value, which, as we have seen, is a constant. 

    \item Regarding tasks $t$, voters of type $T$ schedule them such that each task $t_i$ is completed twice at time $1$, twice at time $2$ and so on. So, regardless of the order of tasks $t$ in $S$, the first task of $S$ is not late for any voter, the second task of $S$ is late by $1$ unit of time for $2$ voters, the third task is late by $1$ unit of time for $2$ voters, by $2$ units of time for two voters and so on. Therefore, the sum of the tardiness of $t$ tasks for voters of type $T$ is also the same for each schedule $S$ in which $t$ tasks precede $dum$ tasks. Let $T_t$ denote this sum of tardiness. 
    
    Voters $D$ schedule all tasks $t$ after $n+1$ except one task $t_j$ (for the $j$-th voter of type $D$), and this task is completed at time $d_j$. Therefore in $S$, each task $t_j$ is always early for all voters $D$ except one, the $j$-th voter of type $D$, and its tardiness for this voter is equal to $\max(0,C_{t_j}(S)-d_j)$.
\end{itemize}
The sum of the tardiness $T(S)$ in schedule $S$ is thus equal to:
$$
T(S)=T_{dum}+T_t+\sum_{t_j \in J_t} \max(0,C_{t_j}(S)-d_j)
$$
Since $T_{dum}$ and $T_t$ do not depend on the order of the tasks in $S$ as long as all tasks $t$ are scheduled first and all tasks $dum$ are scheduled afterwards, the tardiness of schedule $S$ only depends on the position of tasks $t$ relatively to the due dates of the \chainsSigmaT problem. 

We will now prove that there exists a solution $S$ for the instance of the \sigmaTgraphproblem described above such that $T(S) \leq T_{dum}+T_t+K$, if and only if there exists a schedule $S'$ for \chainsSigmaT problem such that the tardiness is smaller than or equal to $K$. In other words, the answer to the question of \sigmaTgraphproblem is then ``yes" if and only if the answer to the question of the corresponding instance of \chainsSigmaT is ``yes". 

Let us assume first that there is a solution $S$ of \sigmaTgraphproblem such that $T(S) \leq T_{dum}+T_t+K$. It means that $\sum_{t_j \in J} \max(0,C_{t_j}(S)-d_j)\leq K$. Let $S'$ be a schedule of tasks of \chainsSigmaT such that the completion time of task $j$ in $S'$ is equal to the completion time of $t_j$ in $S$. We have $\sum_{j \in J} \max(0,C_{j}(S')-d_j)\leq K$, and this solution is feasible since the precedence constraints between the tasks of the \chainsSigmaT problem are the same than between the $t$ tasks. The answer to the question of the \chainsSigmaT is then ``yes".

Let us now assume that there is a feasible solution (schedule) $S'$ of \chainsSigmaT such that the total tardiness is smaller than or equal to $K$. If we create solution $S$ such that the completion time of task $t_j$ in $S$ is equal to the completion time of $j$ in $S'$, we then have $\sum_{t_j \in J_t} \max(0,C_{t_j}(S)-d_j)\leq K$. The $dum$ tasks are then scheduled in any order. Such a solution has then a total tardiness of  $T_t+T_{dum}+K$. This solution is feasible since the precedence constraints between tasks of the \chainsSigmaT problem are the same than between the $t$ tasks. This implies that the answer to the \sigmaTgraphproblem is thus ``yes".

There is a polynomial time reduction from decision problem \chainsSigmaT, which is strongly NP-complete,  to the decision version of our problem \sigmaTgraph. Problem \sigmaTgraph is thus strongly NP-hard. 
\end{proof}



Since, as we have seen before, with unit tasks graphs, and for any profile $P$ and any schedule $S$, the sum of the deviations in $S$ with respect to profile $P$ is equal to twice the sum of the tardiness in $S$, a schedule minimizing the sum of the deviations among schedules which fulfill the precedence constraints will also minimize the sum of the tardiness. Given Proposition~\ref{prop:sigmaT_graph}, we deduce the following corollary. 
\begin{corol}
    The \sigmaDgraph problem is strongly NP-hard, even when the precedence graph consists in chains.
\label{prop:sigmaD_graph}
\end{corol}


We now show that problem \sigmaUgraph, which aims at minimizing the number of late tasks in the returned schedule, with respect to the preferred schedules of the voters, is also a  strongly NP-hard problem. 

\begin{prop}
The \sigmaUgraph problem is strongly NP-hard, even when the precedence graph only consists in chains.
\label{prop:sigmaUgraph}
\end{prop}

We prove this results by doing a polynomial time reduction from the \chainsSigmaU problem. The decision version of this problem is the following one. An instance of this problem is: 
\begin{itemize}
    \item A set $J'=\{1,\dots,{n}\}$ of $n$ unit tasks. Each task $i$ has a  deadline $d_i$. 
    \item A a acyclic precedence graph of $n$ vertices $\{1,\dots,{n}\}$: there is one edge from vertex $i$ to vertex $j$ if task $i$  has to be scheduled before  task $j$. This graph can be only a set of chains between some tasks. 
    \item  An integer $K'$
\end{itemize} 
The aim of optimization problem is to compute a schedule which fulfills the precedence constraints and which minimizes the number of late tasks (i.e. tasks which are completed after their deadlines). 
The question of the corresponding decision problem is the following one: is there a schedule $S$ which fulfills the precedence constraints and in which at most $K'$ tasks are late ? 

~\cite{Garey1976}  have shown that this 
 problem is strongly NP-hard with general precedence constraints, even with unit time tasks. ~\cite{Lenstra1980} have sharpened this result by showing that this problem remains strongly NP-hard, even if the set of precedence constraints is a set of chains.


Without loss of generality we assume that $d_i \leq n$ (tasks with deadlines larger than $n$ will never be late in a schedule of $n$ unit tasks without idle time). 
We create an instance of \sigmaUgraph as follows.

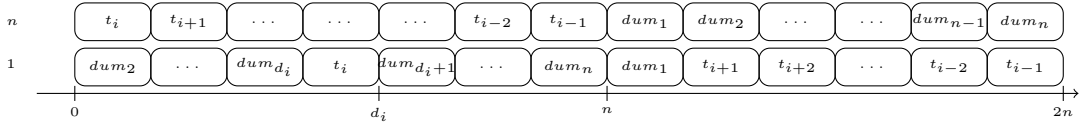
\begin{figure}[H]
\centering
\tiny
\begin{tikzpicture}

\task{$t_i$}{1}{1}{0.6}
\task{$t_{i+1}$}{1}{2}{0.6}
\task{$\dots$}{1}{3}{0.6}
\task{$\dots$}{1}{4}{0.6}
\task{$\dots$}{1}{5}{0.6}
\task{$t_{i-2}$}{1}{6}{0.6}
\task{$t_{i-1}$}{1}{7}{0.6}
\task{$dum_{1}$}{1}{8}{0.6}
\task{$dum_{2}$}{1}{9}{0.6}
\task{$\dots$}{1}{10}{0.6}
\task{$\dots$}{1}{11}{0.6}
\task{$dum_{n-1}$}{1}{12}{0.6}
\task{$dum_{n}$}{1}{13}{0.6}

\task{$dum_2$}{1}{1}{0}
\task{$\dots$}{1}{2}{0}
\task{$dum_{d_i}$}{1}{3}{0}
\task{$t_i$}{1}{4}{0}
\task{$dum_{d_{i}+1}$}{1}{5}{0}
\task{$\dots$}{1}{6}{0}
\task{$dum_n$}{1}{7}{0}
\task{$dum_{1}$}{1}{8}{0}
\task{$t_{i+1}$}{1}{9}{0}
\task{$t_{i+2}$}{1}{10}{0}
\task{$\dots$}{1}{11}{0}
\task{$t_{i-2}$}{1}{12}{0}
\task{$t_{i-1}$}{1}{13}{0}

\draw[->](0-0.5,0)--(13+0.2,0);
\draw (0, 0.1)--(0,-0.1) node[below]{$0$};
\draw (4, 0.1)--(4,-0.1) node[below]{$d_i$};
\draw (7, 0.1)--(7,-0.1) node[below]{$n$};
\draw (13, 0.1)--(13,-0.1) node[below]{$2n$};

\node[text width=1cm] at (-0.4,0.95) {$n$};
\node[text width=1cm] at (-0.4,0.40) {$1$};
\end{tikzpicture}
\caption{Preferred schedules of the $n+1$ voters generated for task $i$. }
\label{fig:voters_sigmaU}
\end{figure}

For each task $i$ of $J'$, we create a task $t_i$ and a task $dum_i$. For each task $i$ we also create $(n+1)$ voters as shown in Figure~\ref{fig:voters_sigmaU}. There are $n$ voters ``of type T" scheduling task $t_i$ first, then $t_{i+1}$ and so forth until $t_n$ and then scheduling tasks $t_1$ to $t_{i-1}$ by increasing index. They then schedule tasks $dum_1$, $dum_2$, \dots $dum_n$. The last voter, ``of type D" schedules task $t_i$ so that it is completed  at time $d_i$, and, if $d_i\neq 1$,  she schedules task $dum_2$ to $dum_{d_i-2}$ by increasing index from time 0 to time $d_i-2$. From time $d_i$, she schedules tasks $dum_{d_i-1}$ to $dum_n$ until time $n$, by increasing index, and she schedules $dum_1$ so that this tasks is completed at time $n+1$. She completes the schedule with tasks $t_{i+1}, \dots, t_n$ by increasing index, followed by tasks $t_1$ to $t_{i-1}$ by increasing index. For any precedence relation between tasks $i$ and $j$ in \chainsSigmaU, we create the same preference relation between tasks $t_i$ and $t_j$ of our \sigmaUgraph instance.

In order to prove Proposition~\ref{prop:sigmaUgraph}, we introduce several lemmas which describe an optimal schedule for the above described instance. As in the proof of Proposition~\ref{prop:sigmaT_graph}, we will see that computing such an optimal solution allow us the associated NP-complete scheduling problem (problem \chainsSigmaU in our case).   

\begin{lemma}
There exists an optimal solution for \sigmaUgraph in which task $dum_1$ completes at time $n+1$.
\label{lem:dum_1}
\end{lemma}
\begin{proof}
All voters schedule $dum_1$ so that it is  completed at time $n+1$. Let $S$ be a schedule in which $dum_i$ does not complete at time $n+1$. We distinguish two sub-cases:
\begin{enumerate}
    \item Task $dum_1$ completes before time  $n+1$: we create schedule $S'$ from $S$ by scheduling $dum_i$ so that it is completed at time $n+1$. We decrease from $1$ unit of time any task scheduled in $S$ between $dum_1$ and time $n+1$. Task $dum_1$ is not late in $S'$ for any voter, just like in $S$ and the task that have been scheduled before cannot become late in $S'$ if they were not in $S$. Therefore the number of late tasks cannot increase from $S$ to $S'$.
    \item Task $dum_1$ completes after $n+1$. We distinguish two sub-cases: 
    \begin{itemize}
        \item If the task $j$ completing at time $n+1$ in $S$ is a $dum$ task,  we create $S'$ from $S$ by swapping the position of $dum_1$ with the task $j$. The unit time penalty for all tasks but $j$ and $dum_1$ are identical between $S$ and $S'$. Task $dum_1$ is in $S'$ on time for the $n(n+1)=n^2+n$ voters, whereas it was late in $S$. On the other hand the unit time cost for task $j$ is increased, but at most by $n^2$ voters, since the $n$ voters of type $D$ already considered it late since they scheduled it before time $n+1$. Overall the unit time penalty is reduced in $S'$ in comparison to $S$.
        \item If the task $j$ completing at time $n+1$ in $S$ is a $t$ task, we create a new schedule $S'$ by scheduling $dum_1$ so that it completes at time $n+1$. We then perform consecutive swaps such that the order on the $t$ tasks is the same in $S$, which is a feasible solution, and $S'$. If there is at least one $t$ task scheduled between $n+1$ and, $C_{dum_1}(S)$, the completion time of $dum_1$ in $S$, we schedule task $j$ at the time slot occupied by the first $t$ task scheduled after $n+1$ in $S$. Let $t_i$ be such a task. This task $t_i$ is then scheduled at the time slot of the following $t$ task which is completed before $C_{dum_1}(S)$, and so on until there is no $t$ task left before $C_{dum_1}(S)$. The final $t$ task moved that way goes on the time slot occupied by $dum_1$ in $S$ (i.e. is completed at time $C_{dum_1}(S)$).  

    \begin{figure}[H]
    \centering
    \small
    \begin{tikzpicture}
        \task{$\dots$}{1}{1}{0}
        \task{$\dots$}{1}{2}{0}
        \task{$\dots$}{1}{3}{0}
        \task{$j$}{1}{4}{0}
        \task{$\dots$}{1}{5}{0}
        \task{$t_i$}{1}{6}{0}
        \task{$dum_a$}{1}{7}{0}
        \task{$dum_b$}{1}{8}{0}
        \task{$t_{i'}$}{1}{9}{0}
        \task{$t_{i''}$}{1}{10}{0}
        \task{$\dots$}{1}{11}{0}
        \task{$dum_1$}{1}{12}{0}
        \task{$\dots$}{1}{13}{0}

        \draw[->](0-0.5,0)--(13+0.2,0);
        \draw (0, 0.1)--(0,-0.1) node[below]{$0$};
        \draw (3, 0.1)--(3,-0.1) node[below]{$n$};
        \draw (13, 0.1)--(13,-0.1) node[below]{$2n$};

        \draw[->]  (3.6,0.7) .. controls (4.5,1) .. (5.4,0.7);
        \draw[->]  (5.6,0.7) .. controls (7,1) .. (8.4,0.7);
        \draw[->]  (8.6,0.7) .. controls (9,0.8) .. (9.4,0.7);
        \draw[->]  (9.6,0.7) .. controls (10.5,1) .. (11.4,0.7);
        
        \draw[->]  (11.4,-0.2) .. controls (7.5,-0.5) .. (3.6,-0.2);

        \node[text width=1cm] at (-0.4,-1.6) {$S'$};
        \node[text width=1cm] at (-0.4,0.40) {$S$};

        \task{$\dots$}{1}{1}{-2}
        \task{$\dots$}{1}{2}{-2}
        \task{$\dots$}{1}{3}{-2}
        \task{$dum_1$}{1}{4}{-2}
        \task{$\dots$}{1}{5}{-2}
        \task{$j$}{1}{6}{-2}
        \task{$dum_a$}{1}{7}{-2}
        \task{$dum_b$}{1}{8}{-2}
        \task{$t_{i}$}{1}{9}{-2}
        \task{$t_{i'}$}{1}{10}{-2}
        \task{$\dots$}{1}{11}{-2}
        \task{$t_{i''}$}{1}{12}{-2}
        \task{$\dots$}{1}{13}{-2}

    \end{tikzpicture}
    \caption{Schedule $S$ and the swaps performed to obtain $S'$.}
    \label{fig:voters_sigmaU_dum1}
    \end{figure}
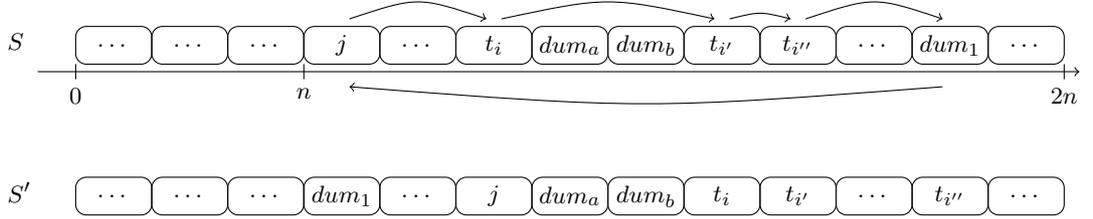

        Note that if the precedence constraints between the $t$ tasks are fulfilled by $S$, they are also fulfilled by $S'$ since the order on the $t$ tasks do not change, just their positions. 
        
        The $t$ tasks which have been moved in $S'$ were considered late in $S$ by all $T$ voters: delaying them do not increase unit time penalty for $T$ voters. Since $D$ voters schedule $t$ task in a cyclic fashion, each $t$ task completes once at time $n+2$, once at time $n+3$ and so on. Therefore delaying a $t$ task by one unit of time between $n+1$ and $2n$ increases its unit time penalty by $1$ (since one additional voter will consider it late). Therefore, when delaying these tasks, the cumulative delay is at most $n$. 
        On the other hand, scheduling $dum_1$ at time $n+1$ decreases the number of late tasks by $n^2+n$ since it is late for all voters in $S$ and on time for all voters in $S'$. This means that the total unit time penalty is smaller in $S'$ than in $S$.
    \end{itemize}
    In all the cases, we managed to generate a solution $S'$ in which $dum_1$ is scheduled between time $n$ and time $n+1$ with the total number of late tasks of $S'$ smaller than or equal to the number of late tasks in $S$. Therefore there always exist an optimal solution in which $dum_1$ is scheduled between $n$ and $n+1$.
\end{enumerate}
\end{proof}


\begin{lemma}
There exists an optimal solution of \sigmaUgraph that fulfills \linebreak  Lemma~\ref{lem:dum_1} and such that there is in this solution a set of successive $dum$ tasks scheduled by increasing index  from time $n$, and none of these tasks are considered late by any voter of type $T$. 
\label{lem:dum_ordre}
\end{lemma}

\begin{proof}
Let $S$ be an optimal solution fulfilling the property of Lemma~\ref{lem:dum_1}: task $dum_1$ is completed at time $n+1$.  


Let us assume that in $S$ some tasks of the $dum$ set starting at time $n$ are not scheduled by increasing index. Let $dum_a$ and $dum_b$ be the two tasks scheduled the earliest in this set and such that $dum_a$ is scheduled before $dum_b$ with $a>b$. Since they are the first two tasks fulfilling this condition any task of this set scheduled before $dum_a$ in $S$ has a smaller index than $a$ and is also scheduled before $dum_a$ in the preferences of voters $T$. 

Let us consider the solution $S'$  obtained from $S$ by swapping the positions of $dum_a$ and $dum_b$. Since $b<a$, $dum_b$ is scheduled before $dum_a$ in the preferences of $T$ voters and since all tasks of the set scheduled before $dum_a$ in $S$ are also scheduled before $dum_a$ in preferences of voters $T$, $dum_a$ cannot be late for voters of type $T$. This means that task $dum_a$ does not become late for $T$ voters in $S'$. This tasks is late for $D$ voters in both $S$ and $S'$ since it is scheduled after $n+1$. 
Since the completion time of $dum_b$ is reduced in $S'$ in comparison to $S$, it cannot be late in $S'$ whereas it was not late in $S$. Therefore, the number of late tasks in $S'$ is not larger than the number of mate tasks in $S$. 

Repeating these swaps until all the $dum$ tasks of the set are scheduled by increasing index, we obtain a new solution in which all of these tasks are on time for $T$ voters and in scheduled by increasing index and such that number of late tasks is not increased in comparison to $S$.
\end{proof}


\begin{lemma}
There exists an optimal solution for \sigmaUgraph which fulfills Lemma~\ref{lem:dum_ordre}, and  such that all $t$ tasks scheduled between time $0$ and time $n$ are scheduled before any $dum$ tasks scheduled between time $0$ and time $n$. 
\label{lem:order_pre_n}
\end{lemma}

\begin{proof}
Let us consider a solution $S$ fulfilling the properties of Lemmas~\ref{lem:dum_1} and~\ref{lem:dum_ordre} 
and such that there is a task $dum_i$ scheduled between time $0$ and time $n$ and such that there is a task $t_j$ scheduled just after $dum_i$ in $S$. Since the task scheduled between time $n$ and $n+1$ is $dum_1$ in $S$, task $t_j$ completes at most at time $n$. 

We create a schedule $S'$ from $S$ by swapping the positions of $dum_i$ and $t_j$. For each date  $k$ between $1$ and $n$, there are $n$ voters of type $T$ scheduling $t_j$ so that it is completed at time $k$. Therefore, advancing $t_j$ by one unit of time between $1$ and $n$, decreases the number of late tasks by $n$. On the other hand, task $dum_i$ is delayed by one unit of time. This does not impact the $T$ voters since they schedule $dum_i$ after time $n+1$. Voters of type $D$ might have an increased unit time penalty for task $dum_i$. Since there are $n$ voters of type $D$, this increases the number of late tasks by at most $n$. Therefore, the number of late tasks in $S'$ is smaller than or equal to the the number of late tasks in $S$.
\end{proof}

\begin{lemma}
There exists an optimal solution for \sigmaUgraph which fulfills Lemma~\ref{lem:order_pre_n}, and in which all the $t$ tasks are scheduled before all $dum$ tasks. Moreover, in this solution, the $dum$ tasks are scheduled in order of increasing indexes. 
\label{lem:t_avant_dum}
\end{lemma}

\begin{proof}
Let $S$ be an optimal solution satisfying the properties of Lemma~\ref{lem:order_pre_n} and such that all tasks $t$ are not scheduled before all $dum$ tasks. Let $dum_i$ be the first $dum$ task to be scheduled in $S$. This implies that $C_{dum_i}\leq n$.  Because of Lemma~\ref{lem:order_pre_n}, task $dum_i$ has to be scheduled after a series of $t$ tasks, and all tasks scheduled after $dum_i$ and before $dum_1$ are $dum$ tasks as well. Let $t_j$ be the first $t$ task scheduled after $dum_i$. As we have seen, $dum_i$ is scheduled after $n+1$ and after a set of $dum$ tasks scheduled by increasing index.


\begin{figure}[H]
    \centering
    \small
    \begin{tikzpicture}
        \task{$t_k$}{1}{1}{0}
        \task{$\dots$}{1}{2}{0}
        \task{$t_l$}{1}{3}{0}
        \task{$dum_i$}{1}{4}{0}
        \task{$\dots$}{1}{5}{0}
        \task{$dum_x$}{1}{6}{0}
        \task{$dum_1$}{1}{7}{0}
        \task{$dum_2$}{1}{8}{0}
        \task{$\dots$}{1}{9}{0}
        \task{$dum_y$}{1}{10}{0}
        \task{$t_j$}{1}{11}{0}
        \task{$\dots$}{1}{12}{0}
        \task{$\dots$}{1}{13}{0}

        \draw[->](0-0.5,0)--(13+0.2,0);
        \draw (0, 0.1)--(0,-0.1) node[below]{$0$};
        \draw (6, 0.1)--(6,-0.1) node[below]{$n$};
        \draw (13, 0.1)--(13,-0.1) node[below]{$2n$};

        \draw[->]  (3.6,0.7) .. controls (7,1) .. (10.4,0.7);
        
        \draw[->]  (10.4,-0.2) .. controls (7,-0.8) .. (3.6,-0.2);

    \end{tikzpicture}
    \caption{Schedule $S$ with the swap performed to obtain $S_{tmp}$. }
    \label{fig:voters_sigmaU_swap}
    \end{figure}
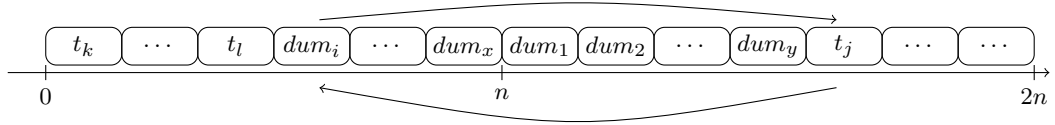

Let $S_{tmp}$ be the schedule obtained by swapping from $S$ the position of $dum_i$ and $t_j$, and let $S'$ be the schedule obtained by swapping from $S$ the position of $dum_i$ and $t_j$ and in which $dum_i$ is re-positioned in the $dum$ set so that the tasks in the set are scheduled by increasing indexes (therefore that $S'$ can also be obtained from $S_{tmp}$ by repositioning $dum_i$ at the right place in the set of $dum$ tasks that follow it in $S$).   

Note that since $S$ fulfills the precedence constraints on the $t$ tasks, then they are fulfilled by $S_{tmp}$ and $S'$ as well since the order on the $t$ tasks does not change. In its  new position in $S_{tmp}$ and $S'$, task $dum_i$ is not late for voters of type $T$ (who schedule $dum_i$ after time $n$). It may be  late for some voters of type $D$ whereas it was not necessarily late for these voters in schedule $S$. Therefore, since there are $n$ voters of type $D$, the number of late tasks due to $dum_i$ is increased by at most $n$ in $S_{tmp}$ and in $S'$. Let us now focus on the total number of late tasks tardiness in $S_{tmp}$. The only tasks whose time slot has changed (compared to is time slot in $S$) is $t_j$, which  was late for all voters of type $T$ in $S$ but now completes at most at time $n$. It thus in $S_{tmp}$ on time for at least $n$ voters of type $T$ (the ones scheduling it between time $n-1$ and $n$). Overall the number of late tasks does not increase in $S_{tmp}$  in comparison to $S$. Since all the voters schedule, in their preferred schedules, the $dum$ tasks by increasing order, the number of late tasks in $S'$ is not larger than the number of late tasks in $S_{tmp}$. Therefore, the number of late tasks does not increase in  $S'$ compared to in  $S$.  

By repeating, if needed, this type of swaps, we obtain an optimal  solution in which all $t$ tasks are scheduled before all $dum$ tasks,  and in which $dum$ tasks are scheduled by increasing indices.
\end{proof}

Starting from any optimal solution $S$ and applying the successive swaps described in Lemmas~\ref{lem:dum_1} to~\ref{lem:t_avant_dum}, we obtain an optimal solution in which tasks $t$ are scheduled first and are followed by $dum$ tasks which are scheduled between time $n$ and $2n$ by increasing indices. Let us now prove Proposition~\ref{prop:sigmaUgraph}.

\begin{proof}
We show that there exists a solution with a total number of late task smaller than or equal to $K'$ for \chainsSigmaU if and only if there exists a solution with a total number of late tasks for \sigmaUgraph smaller than or equal to $K=K'+n(n+1)+\sum_{i=1}^{n} (i-1)n$.

\medskip
Let us first assume that there exists a solution with at most $K$ late tasks for \sigmaUgraph. Thanks to Lemmas~\ref{lem:t_avant_dum}, we know that there exists an optimal solution in which tasks of type $t$ are scheduled before $dum$ tasks, which are scheduled by increasing indices. In such a solution $S$, the number of late tasks can be split into two parts, one independent from the order of the $t$ tasks, and one depending on this order.

Regardless of the order of the $t$ tasks, the $dum$ tasks are all on time for the voters of type $T$, and all (except $dum_1$) late for the voters of type $D$. There are therefore $n-1$ $dum$ tasks late for each of the $n$ voters of type $D$, which amounts to $n(n-1)$ late tasks. Furthermore, $t$ task completes $n$ times at time $1$, $n$ times at time $2$, and so on until time $n$. The $t$ task completing at time $1$ will be on time for all voters of type $T$, the $t$ task completing at time $2$ will be late for $n$ voters of type $T$, the third task will be late for $2n$ voters and so on. This amounts to $\sum_{i=1}^{n}(i-1)n$. For each $i\in\{1,\dots, n \}$, task $t_i$ is on time for $D$ voters,  except for the $i$-th $D$ voter, who scheduled task $t_i$ so that it is completed at time $d_i$.


This means that the total number of late task in $S$ is  $U_t(S)+n(n-1)+\sum_{i=1}^{n} (i-1)n$, where $U_t(S)$ denotes the number of late $t$ tasks in $S$ for voters of type $D$. Since $S$ is an optimal solution and since the answer to the \sigmaUgraph problem is `yes', this means that $U_t(S)\leq K'$.

We label the $t$ tasks according to their position in schedule $S$, which can be described as follows: $t_{S(1)},t_{S(2)},\dots,t_{S(n)},dum_1,dum_2,\dots,dum_n$,  where $S(i)$ denotes the index of the task scheduled in position $i$ in $S$. We consider the schedule $S'$ of tasks of \chainsSigmaU: ${S(1)},{S(2)},\dots,{S(n)}$. 
Note that since $S$ is a feasible solution of \sigmaUgraph and since the precedence constraints on $t$ tasks are the same than on the tasks of the \chainsSigmaU instance, $S'$ is a feasible solution of \chainsSigmaU. In $S'$, task ${S(i)}$ is completed at the same time than task $t_{S(i)}$ in $S$, therefore task ${S(i)}$ is late if and only if ${S(i)}$ is late for the voter scheduling ${S(i)}$ at time $d_{S(i)}$. Therefore if $U_t(S)\leq K'$, the total number of late tasks in $S'$ is also smaller than or equal to $K'$, which means that the answer to the \chainsSigmaU problem is also `yes'. 

\medskip

Reciprocally, if the answer to the \chainsSigmaU problem is `yes', then there exists a schedule $S'$ for \chainsSigmaU such that the total number of late tasks in $S'$ is smaller than or equal to $K'$. We consider $S$ the schedule $t_{S'(1)},\dots,t_{S'(n)}, dum_1, dum_2, \dots, dum_n$ for the \sigmaUgraph problem. Schedule $S$ fulfills the precedence constraints of the \sigmaUgraph instance since these precedence constraints are the same than the precedence constraints on the $t$ tasks of the corresponding instance of \sigmaUgraph. 

Since $S$ fulfills the property of Lemma~\ref{lem:t_avant_dum}, there is a constant number of late task $n(n-1)+\sum_{i=1}^{n}$ for voters of type $T$. The number of late $t$ tasks for voters of type $D$ depends on whether task $t_i$ is scheduled before or after time $d_i$ since only one $D$ voter schedules task $t_i$ before time $n$ (she schedules $t_i$ between times $d_i-1$ and $d_i$). Task $t_i$ is completed in $S$ at the same time than task $i$ in $S'$. Therefore task $t_i$ completes after $d_i$ in $S$ if and only if task $i$ is late in $S'$. Therefore the number of late $t$ tasks in $S$ for voters of type $D$ is equal to the number of late tasks in $S'$. Since the number of late tasks in $S'$ is smaller than or equal to $K'$, the total number of late tasks in $S$ is smaller than or equal to $K'+n(n+1)+\sum_{i=1}^{n} (i-1)n$ and the answer to  \sigmaUgraph is then `yes'.


The answer to the \sigmaUgraph problem is `yes' if and only if the answer to the \chainsSigmaU problem is `yes'. Since the decision version of problem  \chainsSigmaU is strongly NP-complete~\citep{Lenstra1980}, we conclude that the decision version or problem \sigmaUgraph is also strongly NP-complete.  
\end{proof}

Proposition~\ref{prop:sigmaT_graph} shows that problem \sigmaTgraph is strongly NP-hard, while Proposition~\ref{prop:sigmaUgraph} shows that problem \sigmaUgraph is strongly NP-hard, even if the precedence graphs are only made of chains of tasks. Since, as we have seen in Section~\ref{subsec:genralization}, problem  \sigmaT is a special case of the Distance Criterion, and problem  \sigmaU is a special case of the Binary Criterion, we get the following corollary. 

\begin{corol}
    Returning an optimal solution for the Distance Criterion or the Binary Criterion are strongly NP-hard problems when there are imposed precedence constraints. This is true even with precedence graphs only made of chains.  
\label{corol:distanceetbinary}
\end{corol} 


\section{Conclusion}
\label{sec:conclusion}
In this paper, we studied the collective scheduling problem with unit size tasks, which can also be seen as a  collective ranking problem since tasks of length 1 can be considered as items and preferred schedules as preferred rankings. 

We introduced two general objective functions, one based on a distance, and the other one on a binary criterion. The distance based function minimizes the average distance between  the returned schedule (or ranking) and the preferences of the voters (expressed as preferred schedules or preferred intervals for each task). It generalizes already known rules that minimize of the average deviation (\sigmaD), or the average tardiness (\sigmaT). The binary function generalizes the rule \sigmaU that minimizes the average number of late tasks. These rules can be applied in polynomial time even if we add release dates and deadlines constraints on the tasks.  

We studied these two general rules from an axiomatic point of view when we infer release dates and deadlines from the preferences of the voters, showing that they do not fulfills release date or deadline consistency, but that  they fulfill temporal unanimity, three axioms that we have introduced in this paper.

We have also shown that the rule EMD which schedules the tasks by increasing median completion time (or by increasing median place in a ranking if we consider rankings instead of schedules), is a 2-approximation for the sum of the deviation (or the sum of the tardiness) minimization. Note that the rule which minimizes the sum of the deviations between a collective ranking and a set of preferred ranking, is known as Spearman's rule: interestingly, EMD is thus a 2-approximation of Spearman's rule.

Last but not least, we studied the case where there are precedence constraints between the tasks. Note that precedence constraints also  make sense in the context of rankings if items, a constraint between two items $a$ and $b$ saying that item $a$ has to be ranked higher than item $b$. We showed that if the precedence constraints are fulfilled by the preferred schedules (or rankings) of the voters, then it is easy to get an optimal schedule (ranking) which fulfills the precedence constraints while minimizing the average deviation (or the average tardiness). When the preferred schedules do not necessarily fulfill the constraints, we showed that on the contrary, it is NP-hard to find a schedule that fulfills the precedence constraints while minimizing the average deviation (or the average tardiness, or the average number of late tasks).

Whether the minimization of the number of late tasks is an NP-hard problem with inferred precedence constraints,  remains an open question. Looking for exact but efficient,  or approximate, algorithms to compute solutions for the above mentioned NP-hard problems is also an interesting research direction.  Finally, in context in which release date or deadline consistency are important axioms, finding a rule which fulfills these axioms while returning good solutions with respect to the Distance or the Binary criterion would also be a promising research direction.

\bibliography{bibliography_cs_unit}

\end{document}